\documentclass[trackchanges, twocolumn, twocolappendix]{aastex7}
\usepackage{amsmath}
\usepackage{xcolor}

\usepackage{xcolor}

\begin{document}

\title{Searching for Exotrojans in Pulsar Systems}

\author[0009-0006-8984-9220]{Jackson D. Taylor}
\affiliation{Department of Physics and Astronomy, West Virginia University, P.O. Box 6315,
Morgantown, WV 26506, USA}
\affiliation{Center for Gravitational Waves and Cosmology, West Virginia University, Chestnut Ridge Research Building, Morgantown, WV 26505, USA}
\email{jdt00012@mix.wvu.edu, jacksondtaylor4@gmail.com}

\author[0000-0001-8384-5049]{Emmanuel Fonseca}
\affiliation{Department of Physics and Astronomy, West Virginia University, P.O. Box 6315,
Morgantown, WV 26506, USA}
\affiliation{Center for Gravitational Waves and Cosmology, West Virginia University, Chestnut Ridge Research Building, Morgantown, WV 26505, USA}
\email{emmanuel.fonseca@mail.wvu.edu}

\author[0000-0002-2554-0674]{Lankeswar Dey}
\affiliation{Department of Physics and Astronomy, West Virginia University, P.O. Box 6315,
Morgantown, WV 26506, USA}
\affiliation{Center for Gravitational Waves and Cosmology, West Virginia University, Chestnut Ridge Research Building, Morgantown, WV 26505, USA}
\email{lankeswar.dey@nanograv.org}

\author[0000-0003-2526-2683]{Sergey Zharikov}
\affiliation{Instituto de Astronom\'{i}a, Universidad Nacional Aut\'{o}noma de M\'{e}xico,  Apdo. Postal 877, Ensenada, Baja California 22800, M\'{e}xico}
\email{zhar@astro.unam.mx}

\author[0000-0002-8139-8414]{Aida Kirichenko}
\affiliation{Instituto de Astronom\'{i}a, Universidad Nacional Aut\'{o}noma de M\'{e}xico, Apdo. Postal 877, Ensenada, Baja California 22800, M\'{e}xico}
\email{aida@astro.unam.mx}

\author[0000-0003-4090-9780]{Joseph Glaser}
\affiliation{Department of Physics and Astronomy, West Virginia University, P.O. Box 6315, Morgantown, WV 26506, USA}
\affiliation{Center for Gravitational Waves and Cosmology, West Virginia University, Chestnut Ridge Research Building, Morgantown, WV 26505, USA}
\email{joseph.glaser@mail.wvu.edu}

\author[0000-0001-5134-3925]{Gabriella Agazie}
\affiliation{Center for Gravitation, Cosmology and Astrophysics, Department of Physics and Astronomy, University of Wisconsin-Milwaukee,\\ P.O. Box 413, Milwaukee, WI 53201, USA}
\email{gabriella.agazie@nanograv.org}
\author[0000-0002-8935-9882]{Akash Anumarlapudi}
\affiliation{Department of Physics and Astronomy, University of North Carolina, Chapel Hill, NC 27599, USA}
\email{akasha@unc.edu}
\author[0000-0003-0638-3340]{Anne M. Archibald}
\affiliation{Newcastle University, NE1 7RU, UK}
\email{anne.archibald@nanograv.org}
\author[0009-0008-6187-8753]{Zaven Arzoumanian}
\affiliation{X-Ray Astrophysics Laboratory, NASA Goddard Space Flight Center, Code 662, Greenbelt, MD 20771, USA}
\email{zaven.arzoumanian@nanograv.org}
\author[0000-0003-2745-753X]{Paul T. Baker}
\affiliation{Department of Physics and Astronomy, Widener University, One University Place, Chester, PA 19013, USA}
\email{paul.baker@nanograv.org}
\author[0000-0003-3053-6538]{Paul R. Brook}
\affiliation{Institute for Gravitational Wave Astronomy and School of Physics and Astronomy, University of Birmingham, Edgbaston, Birmingham B15 2TT, UK}
\email{paul.brook@nanograv.org}
\author[0000-0002-6039-692X]{H. Thankful Cromartie}
\affiliation{National Research Council Research Associate, National Academy of Sciences, Washington, DC 20001, USA resident at Naval Research Laboratory, Washington, DC 20375, USA}
\email{thankful.cromartie@nanograv.org}
\author[0000-0002-1529-5169]{Kathryn Crowter}
\affiliation{Department of Physics and Astronomy, University of British Columbia, 6224 Agricultural Road, Vancouver, BC V6T 1Z1, Canada}
\email{kathryn.crowter@nanograv.org}
\author[0000-0002-2185-1790]{Megan E. DeCesar}
\altaffiliation{Resident at the Naval Research Laboratory}
\affiliation{Department of Physics and Astronomy, George Mason University, Fairfax, VA 22030, resident at the U.S. Naval Research Laboratory, Washington, DC 20375, USA}
\email{megan.decesar@nanograv.org}
\author[0000-0002-6664-965X]{Paul B. Demorest}
\affiliation{National Radio Astronomy Observatory, 1003 Lopezville Rd., Socorro, NM 87801, USA}
\email{paul.demorest@nanograv.org}
\author[0000-0001-8885-6388]{Timothy Dolch}
\affiliation{Department of Physics, Hillsdale College, 33 E. College Street, Hillsdale, MI 49242, USA}
\affiliation{Eureka Scientific, 2452 Delmer Street, Suite 100, Oakland, CA 94602-3017, USA}
\email{timothy.dolch@nanograv.org}
\author[0000-0001-7828-7708]{Elizabeth C. Ferrara}
\affiliation{Department of Astronomy, University of Maryland, College Park, MD 20742, USA}
\affiliation{Center for Research and Exploration in Space Science and Technology, NASA/GSFC, Greenbelt, MD 20771}
\affiliation{NASA Goddard Space Flight Center, Greenbelt, MD 20771, USA}
\email{elizabeth.ferrara@nanograv.org}
\author[0000-0001-5645-5336]{William Fiore}
\affiliation{Department of Physics and Astronomy, University of British Columbia, 6224 Agricultural Road, Vancouver, BC V6T 1Z1, Canada}
\email{william.fiore@nanograv.org}
\author[0000-0001-7624-4616]{Gabriel E. Freedman}
\affiliation{NASA Goddard Space Flight Center, Greenbelt, MD 20771, USA}
\email{gabriel.freedman@nanograv.org}
\author[0000-0001-6166-9646]{Nate Garver-Daniels}
\affiliation{Department of Physics and Astronomy, West Virginia University, P.O. Box 6315, Morgantown, WV 26506, USA}
\affiliation{Center for Gravitational Waves and Cosmology, West Virginia University, Chestnut Ridge Research Building, Morgantown, WV 26505, USA}
\email{nathaniel.garver-daniels@nanograv.org}
\author[0000-0001-8158-683X]{Peter A. Gentile}
\affiliation{Department of Physics and Astronomy, West Virginia University, P.O. Box 6315, Morgantown, WV 26506, USA}
\affiliation{Center for Gravitational Waves and Cosmology, West Virginia University, Chestnut Ridge Research Building, Morgantown, WV 26505, USA}
\email{peter.gentile@nanograv.org}
\author[0000-0003-1884-348X]{Deborah C. Good}
\affiliation{Department of Physics and Astronomy, University of Montana, 32 Campus Drive, Missoula, MT 59812}
\email{deborah.good@nanograv.org}
\author[0000-0003-2742-3321]{Jeffrey S. Hazboun}
\affiliation{Department of Physics, Oregon State University, Corvallis, OR 97331, USA}
\email{jeffrey.hazboun@nanograv.org}
\author[0000-0003-1082-2342]{Ross J. Jennings}
\altaffiliation{NANOGrav Physics Frontiers Center Postdoctoral Fellow}
\affiliation{Department of Physics and Astronomy, West Virginia University, P.O. Box 6315, Morgantown, WV 26506, USA}
\affiliation{Center for Gravitational Waves and Cosmology, West Virginia University, Chestnut Ridge Research Building, Morgantown, WV 26505, USA}
\email{ross.jennings@nanograv.org}
\author[0000-0001-6607-3710]{Megan L. Jones}
\affiliation{Center for Gravitation, Cosmology and Astrophysics, Department of Physics and Astronomy, University of Wisconsin-Milwaukee,\\ P.O. Box 413, Milwaukee, WI 53201, USA}
\email{megan.jones@nanograv.org}
\author[0000-0001-6295-2881]{David L. Kaplan}
\affiliation{Center for Gravitation, Cosmology and Astrophysics, Department of Physics and Astronomy, University of Wisconsin-Milwaukee,\\ P.O. Box 413, Milwaukee, WI 53201, USA}
\email{kaplan@uwm.edu}
\author[0000-0002-0893-4073]{Matthew Kerr}
\affiliation{Space Science Division, Naval Research Laboratory, Washington, DC 20375-5352, USA}
\email{matthew.kerr@nanograv.org}
\author[0000-0003-0721-651X]{Michael T. Lam}
\affiliation{SETI Institute, 339 N Bernardo Ave Suite 200, Mountain View, CA 94043, USA}
\affiliation{School of Physics and Astronomy, Rochester Institute of Technology, Rochester, NY 14623, USA}
\affiliation{Laboratory for Multiwavelength Astrophysics, Rochester Institute of Technology, Rochester, NY 14623, USA}
\email{michael.lam@nanograv.org}

\author{T. Joseph W. Lazio}
\affiliation{Independent Researcher}
\email{joseph.lazio@nanograv.org}

\author[0000-0003-1301-966X]{Duncan R. Lorimer}
\affiliation{Department of Physics and Astronomy, West Virginia University, P.O. Box 6315, Morgantown, WV 26506, USA}
\affiliation{Center for Gravitational Waves and Cosmology, West Virginia University, Chestnut Ridge Research Building, Morgantown, WV 26505, USA}
\email{duncan.lorimer@nanograv.org}
\author[0000-0001-5373-5914]{Jing Luo}
\altaffiliation{Deceased}
\affiliation{Department of Astronomy \& Astrophysics, University of Toronto, 50 Saint George Street, Toronto, ON M5S 3H4, Canada}
\email{jing.luo@nanograv.org}
\author[0000-0001-5229-7430]{Ryan S. Lynch}
\affiliation{Green Bank Observatory, P.O. Box 2, Green Bank, WV 24944, USA}
\email{ryan.lynch@nanograv.org}
\author[0000-0001-5481-7559]{Alexander McEwen}
\affiliation{Center for Gravitation, Cosmology and Astrophysics, Department of Physics and Astronomy, University of Wisconsin-Milwaukee,\\ P.O. Box 413, Milwaukee, WI 53201, USA}
\email{alexander.mcewen@nanograv.org}
\author[0000-0001-7697-7422]{Maura A. McLaughlin}
\affiliation{Department of Physics and Astronomy, West Virginia University, P.O. Box 6315, Morgantown, WV 26506, USA}
\affiliation{Center for Gravitational Waves and Cosmology, West Virginia University, Chestnut Ridge Research Building, Morgantown, WV 26505, USA}
\email{maura.mclaughlin@nanograv.org}
\author[0000-0002-4642-1260]{Natasha McMann}
\affiliation{Department of Physics and Astronomy, Vanderbilt University, 2301 Vanderbilt Place, Nashville, TN 37235, USA}
\email{natasha.mcmann@nanograv.org}
\author[0000-0001-8845-1225]{Bradley W. Meyers}
\affiliation{Australian SKA Regional Centre (AusSRC), Curtin University, Bentley, WA 6102, Australia}
\affiliation{International Centre for Radio Astronomy Research (ICRAR), Curtin University, Bentley, WA 6102, Australia}
\email{bradley.meyers@nanograv.org}
\author[0000-0002-3616-5160]{Cherry Ng}
\affiliation{Dunlap Institute for Astronomy and Astrophysics, University of Toronto, 50 St. George St., Toronto, ON M5S 3H4, Canada}
\email{cherry.ng@nanograv.org}
\author[0000-0002-6709-2566]{David J. Nice}
\affiliation{Department of Physics, Lafayette College, Easton, PA 18042, USA}
\email{niced@lafayette.edu}
\author[0000-0001-5465-2889]{Timothy T. Pennucci}
\affiliation{Institute of Physics and Astronomy, E\"{o}tv\"{o}s Lor\'{a}nd University, P\'{a}zm\'{a}ny P. s. 1/A, 1117 Budapest, Hungary}
\email{timothy.pennucci@nanograv.org}
\author[0000-0002-8509-5947]{Benetge B. P. Perera}
\affiliation{Arecibo Observatory, HC3 Box 53995, Arecibo, PR 00612, USA}
\email{benetge.perera@nanograv.org}
\author[0000-0002-8826-1285]{Nihan S. Pol}
\affiliation{Department of Physics, Texas Tech University, Box 41051, Lubbock, TX 79409, USA}
\email{nihan.pol@nanograv.org}
\author[0000-0002-2074-4360]{Henri A. Radovan}
\affiliation{Department of Physics, University of Puerto Rico, Mayag\"{u}ez, PR 00681, USA}
\email{henri.radovan@nanograv.org}
\author[0000-0001-5799-9714]{Scott M. Ransom}
\affiliation{National Radio Astronomy Observatory, 520 Edgemont Road, Charlottesville, VA 22903, USA}
\email{sransom@nrao.edu}
\author[0000-0002-5297-5278]{Paul S. Ray}
\affiliation{Space Science Division, Naval Research Laboratory, Washington, DC 20375-5352, USA}
\email{paul.ray@nanograv.org}
\author[0000-0003-4391-936X]{Ann Schmiedekamp}
\affiliation{Department of Physics, Penn State Abington, Abington, PA 19001, USA}
\email{ann.schmiedekamp@nanograv.org}
\author[0000-0002-1283-2184]{Carl Schmiedekamp}
\affiliation{Department of Physics, Penn State Abington, Abington, PA 19001, USA}
\email{carl.schmiedekamp@nanograv.org}
\author[0000-0002-7283-1124]{Brent J. Shapiro-Albert}
\affiliation{Department of Physics and Astronomy, West Virginia University, P.O. Box 6315, Morgantown, WV 26506, USA}
\affiliation{Center for Gravitational Waves and Cosmology, West Virginia University, Chestnut Ridge Research Building, Morgantown, WV 26505, USA}
\affiliation{Giant Army, 915A 17th Ave, Seattle WA 98122}
\email{brent.shapiro-albert@nanograv.org}
\author[0000-0001-9784-8670]{Ingrid H. Stairs}
\affiliation{Department of Physics and Astronomy, University of British Columbia, 6224 Agricultural Road, Vancouver, BC V6T 1Z1, Canada}
\email{stairs@astro.ubc.ca}
\author[0000-0002-7261-594X]{Kevin Stovall}
\affiliation{National Radio Astronomy Observatory, 1003 Lopezville Rd., Socorro, NM 87801, USA}
\email{kevin.stovall@nanograv.org}
\author[0000-0002-2820-0931]{Abhimanyu Susobhanan}
\affiliation{Max-Planck-Institut f{\"u}r Gravitationsphysik (Albert-Einstein-Institut), Callinstra{\ss}e 38, D-30167 Hannover, Germany\\Leibniz Universit{\"a}t Hannover, D-30167 Hannover, Germany}
\email{abhimanyu.susobhanan@nanograv.org}
\author[0000-0002-1075-3837]{Joseph K. Swiggum}
\altaffiliation{NANOGrav Physics Frontiers Center Postdoctoral Fellow}
\affiliation{Department of Physics, Lafayette College, Easton, PA 18042, USA}
\email{joseph.swiggum@nanograv.org}
\author[0000-0001-9678-0299]{Haley M. Wahl}
\affiliation{Department of Physics and Astronomy, West Virginia University, P.O. Box 6315, Morgantown, WV 26506, USA}
\affiliation{Center for Gravitational Waves and Cosmology, West Virginia University, Chestnut Ridge Research Building, Morgantown, WV 26505, USA}
\email{haley.wahl@nanograv.org}

%%%

% \collaboration{1000}{(NANOGrav collaboration)}
\correspondingauthor{Jackson D. Taylor}
\email{jdt00012@mix.wvu.edu, jacksondtaylor4@gmail.com}

\begin{abstract}

Trojan asteroids are found in the equilateral triangle Lagrange points of the Sun-Jupiter system in great number, though they also exist less prolifically in other parts of the Solar System. Despite up to planetary mass Trojans being predicted in extrasolar systems (i.e. exotrojans), they remain unconfirmed, although strong candidate evidence has emerged recently.
For the first time, we extend the search for exotrojans to radio pulsars with low-mass ($\sim0.01\,\rm{M}_\odot$) companions using accurately measured pulse times of arrival.
% For the first time, we extend the search for exotrojans to binary radio pulsars---pulsars orbiting any type of companion---using accurately measured pulse times of arrival.
% We offer initial observational limits while their possible existence remains uncertain.
With techniques developed for detecting the reflex motion of a star due to a librating Trojan, we place $\sim 1\,\rm{M}_\oplus$ upper mass constraints on potential exotrojans around eight pulsars observed in the NANOGrav 15-year data set. 
% With techniques developed for detecting the reflex motion of a star due to a librating Trojan, we place \textcolor{red}{$\sim 1\,\rm{M}_\oplus$} upper mass constraints on potential exotrojans around eight pulsars in binary systems with low-mass ($\sim0.01\,\rm{M}_\odot$) companions observed in the NANOGrav 15-year data set. 
% With techniques developed for detecting the reflex motion of a star due to a librating Trojan, we place reasonable upper mass constraints ($\sim 1\,\rm{M}_\oplus$) on potential exotrojans around eight pulsars in binary systems with low-mass ($\sim0.01\,\rm{M}_\odot$) companions observed in the NANOGrav 15-year data set. 
We find weak evidence consistent with $\sim2$--4$\,\rm{M}_{\rm J}$ exotrojans in the PSR~J0023+0923 and PSR~J1705$-$1903 binary systems, though the signals likely have a different, unknown source. 
We also place a libration-independent upper mass constraint of $\sim8$\,M$_{\rm J}$ on exotrojans in the PSR~J1641+8049 system by looking for an inconsistency between the times of superior conjunction as measured by optical light curves and those predicted by radio timing.
These results offer initial observational constraints on the existence of exotrojans around pulsars, while their possible formation mechanisms remain unexplored.

\end{abstract}



\section{Introduction} 
\label{sec:intro}

% Trojans are objects that reside in the $L_4$ or $L_5$ Lagrange points of an otherwise two-body system. 
Trojans are objects that reside in the $L_4$ or $L_5$ Lagrange points of two massive bodies. 
% These so-called co-orbitals share the same mean motion as the more-massive companion (known as the dominant companion) around its higher-mass host, and they appear to trail or lead the companion by $60^\circ$ throughout its orbit. 
In our case, the relevant configuration consists of a pulsar, its low-mass companion (which we call the secondary), and a possible Trojan. The Trojan shares the secondary's mean motion and trails or leads the secondary by $60^\circ$ throughout its orbit.
In the Jupiter-Sun system alone, there are over ten thousand confirmed Trojans \citep[e.g.,][]{ten_thous} along with estimates of hundreds of thousands more that have diameters above $\sim$ 2\,km \citep[see, e.g.,][]{hundred_thous_2000, yoshida_hundred_thous_2005}. Every Solar System planet currently has Trojans of its own, barring Mercury---which may lack confirmed Trojans due to observational difficulties---and there are even Trojans of the Saturn-Tethys system. 
 
% No natural satellites exist in the $L_1$, $L_2$, or $L_3$ Lagrange-point configurations because these are only semi-stable points. The triangular $L_4$ and $L_5$ points are locally stable, meaning they are stable provided the Trojan does not stray too far from the equilibrium point (see \citealt{danby1964stability, giuppone2010dynamics, leleu_mainpaper_2015} for discussions on the size of the stability region). Furthermore, in the case of circular orbits, the triangular Lagrange points are only stable if 
% \begin{equation}
% \label{eq:gascheau_stability}
%     \frac{m_{\mathrm{p}} m_{\mathrm{c}} + m_{\mathrm{p}} m_{\mathrm{T}} + m_{\mathrm{c}} m_{\mathrm{T}}}{(m_{\mathrm{p}} + m_{\mathrm{c}} + m_{\mathrm{T}})^2} < \frac{1}{27},
% \end{equation}

% where $m_{\rm p}$ is the mass of the host star (a pulsar in this study), $m_{\rm c}$ is the mass of the star's dominant companion, and $m_{\rm T}$ is the mass of the Trojan \citep[][]{gascheau1843}. \citet{danby1964stability} showed that introducing a non-zero orbital eccentricity generally decreases the window of stable masses.

% There is good reason to expect exotrojans to exist besides the Trojan abundance in our Solar System. 
The abundance of Trojans in our Solar System suggests that extrasolar Trojans (exotrojans) should also be common.
Indeed, any robust Solar System formation theory must allow for their creation, and many of these theories do not just predict Trojan asteroids but also, surprisingly, Trojans of planetary mass \citep[e.g.,][]{laughlin2002extrasolar, cresswell2008three, lyra2009_trojanearthformation, montesinos_dust_accumulation_trojanorigin_2020}. 
% Once Trojans form, they can exist stably provided the Trojan does not stray too far from the equilibrium point (see \citealt{danby1964stability, giuppone2010dynamics, leleu_mainpaper_2015} for discussions on the size of the stability region). 
Once formed, it has been shown that these Trojans can find themselves in relatively large stable regions  \citep{danby1964stability, giuppone2010dynamics, leleu_mainpaper_2015}.
In the case of circular orbits, the triangular Lagrange points are only stable if 
\begin{equation}
\label{eq:gascheau_stability}
    \frac{m_{\mathrm{p}} m_{\mathrm{c}} + m_{\mathrm{p}} m_{\mathrm{T}} + m_{\mathrm{c}} m_{\mathrm{T}}}{(m_{\mathrm{p}} + m_{\mathrm{c}} + m_{\mathrm{T}})^2} < \frac{1}{27},
\end{equation}
where $m_{\rm p}$ is the mass of the host star (a pulsar in this study), $m_{\rm c}$ is the mass of the secondary (the pulsar's dominant companion), and $m_{\rm T}$ is the mass of the Trojan (the pulsar's lowest mass companion; \citealt{gascheau1843}). \citet{danby1964stability} showed that introducing a non-zero orbital eccentricity generally decreases the window of stable masses.

% Despite their prevalence in our Solar System and their accompanying formation models, no Trojans outside the Solar System have been confirmed, though a recent study by \citet{balsalobre-ruza_troy3_2024}, which continued the works of \citet{lillo-box_troy1_2018, lillo-box_troy2_2018}, identified one of the strongest candidates to date with a \mbox{$3$-$\sigma$} detection. 
Unless the Trojan is librating around its stable point, the measured gravitational effects on the star from a single orbiting companion are degenerate with those of two co-orbiting companions up to $\mathcal{O}(e_{\rm{o}}^2)$ \citep[][]{leleu_theory_alphamethod_2017},
where $e_{\rm o}$ is the orbital eccentricity. 
This degeneracy helps explain why, despite their prevalence in our Solar System and their accompanying formation models, no Trojans outside the Solar System have yet been confirmed. 
\citet{ford_observational_2006} showed that combining radial velocity data with transit-timing data breaks this degeneracy. \citet{leleu_theory_alphamethod_2017} further formalized their approach into the robust \mbox{$\alpha$-test} method to include the results of several transits. \citet{balsalobre-ruza_troy3_2024} applied this technique to identify one of the strongest candidates to date with a \mbox{$3$-$\sigma$} detection, building on the earlier works of \citet{lillo-box_troy1_2018, lillo-box_troy2_2018}. 
% These studies used the \mbox{$\alpha$-test} method as developed by \citep[][]{leleu_theory_alphamethod_2017}, which exploits any Trojan-caused inconsistencies between the transit-timing and radial velocity methods---discrepancies first identified by \citeauthor{ford_observational_2006} (\citeyear{ford_observational_2006}; see Section~\ref{sec:sup_conj} of this work). 
These works are part of, or helped form the foundation for, the broader TROY project,\footnote{\url{www.troy-project.com}} which seeks to discover the first extrasolar Trojan planets. While most of their work has been confined to optical observations of main-sequence stars, \cite{balsalobre-ruza_tentative_radioALMA_2023} potentially identified an accumulation of gas and dust that might be a forming Trojan in the PDS~70 protoplanetary disk \citep[e.g.,][]{keppler2018discovery, haffert_pds70_2019}. 

Other studies have searched for photometric transits of the Trojans themselves. \citet{janson_kepler_indivdual2013} used the \textit{Kepler} space telescope to rule out transiting Trojans down to 1\,R$_\oplus$ in 2244 systems, and \citet{moldovan_MOST_2010searching} used the \textit{MOST} satellite to place a lunar mass upper limit on a Trojan in the HD~209458 system. Rather than looking for transits from individual Trojans, \citet{hippke2015statistical} looked for a cumulative Trojan effect across nearly 4000 planets by adding their \textit{Kepler} light curves coherently by orbital phase, and identified potential transits of the stacked light curves at the expected $L_4$ and $L_5$ orbital phases. 
The main problem with identifying Trojans transiting their host stars is the fact that we expect exotrojans to have a broad range of orbital inclinations, as is the case for the Jupiter Trojans \citep[][]{yoshida_hundred_thous_2005, nesvorny_distributions_2013, bottke2023origin}. 
Therefore, the dominant \textit{Kepler} planet transiting its host star does not guarantee that any Trojan planets will also transit during any given observation. In contrast, the \mbox{$\alpha$-test} method does not rely on in-plane Trojans.

% There is good reason to expect exotrojans to exist besides their abundance in our Solar System. Indeed, any robust Solar System formation theory must allow for their creation, and many of these theories do not just predict Trojan asteroids but also Trojans of planetary mass \citep[e.g.,][]{laughlin2002extrasolar, cresswell2008three, lyra2009_trojanearthformation, montesinos_dust_accumulation_trojanorigin_2020}. 
While almost every exotrojan search has been done around main-sequence stars, as detailed above, no study has looked for them around pulsar binary systems. 
% In particular, Trojans may reside in black widow pulsar systems---binaries characterized by a millisecond pulsar (MSP) with a low-mass ($\sim0.01\,\mathrm{M_\odot}$) companion \citep[e.g.,][]{shahbaz2017black_widow}---and are therefore likely to follow the stability criterion given in Equation~\ref{eq:gascheau_stability}. 
Just as we expect Trojans to form around ordinary stars during the protoplanetary disk phase \citep[e.g.,][]{montesinos_dust_accumulation_trojanorigin_2020}, Trojans may also form around pulsars when a disk is present and then settle via gas drag \citep[][]{lyra2009_trojanearthformation} into stable so-called tadpole (T-type) orbits. Such a disk can form in at least two possible ways.
% : supernova (SN) fallback disks or accretion disks during the transition-to-MSP phase. 
One type of pulsar disk, the supernova fallback disk, forms when some of the material from the supernova that formed the pulsar returns to the pulsar and collapses into a disk \citep[][]{behrens_nanograv_2020, xu2024disk}. Such disks are a plausible formation channel for pulsar planets \citep[][]{lin1991SNdiskformation, wolszczan1992_firstexoplanet}, and would explain why the planets around PSR~B1257+12 have similar inclinations \citep[][]{konacki2003_pulsarplanets_plane}.

The other possible type of pulsar disk
% , the accretion disk that turns a canonical pulsar \citep[e.g.,][]{condon2016essential} into an MSP (see \citet{papitto2021transitionalMSP} and references therein),
forms when the compact pulsar in a binary system strips material from its evolving, ablated companion. This material forms an accretion disk, which collects onto the pulsar and transitions the pulsar into a millisecond pulsar (MSP) \citep[][]{archibald2009radio_xraybinarylink, papitto2021transitionalMSP}. Before the accretion disk dissipates, it may form planets \citep[][]{patruno2017third_gen, behrens_nanograv_2020}, an asteroid belt \citep[][]{shannon2013asteroidbelt}, or Trojan planets in 1:1 resonance with the pulsar's companion, as in the case of protoplanetary-disk-formed Trojans \citep[][]{lyra2009_trojanearthformation, montesinos_dust_accumulation_trojanorigin_2020}. 
Potential Trojans, too, may be stripped of their material, but the timescale of complete dissipation is unclear, with some estimates ranging up to a few Gyr for the secondary \citep[e.g.][]{ginzburg2020dissipation_timescale}. This picture is complicated by the fact that material from the secondary could continue to fall into the Triangular Lagrange points.
In any case, detecting an interloping Trojan may be possible even if it lasts just a few hundred Myr before complete dissipation.
% In any case, detecting a Trojan may be possible even if it is just a few hundred Myr interloper.
% types of disk: SN fallback disk, or an accretion disk from its companion during the transition to MSP (tMSP) phase

The above pulsar Trojan formation discussion lays out plausible channels, but the possible existence of pulsar Trojans remains speculative. Future theoretical work is required to draw definitive connections between proto-planetary disks and pulsar-centered disks with respect to Trojan creation. 
% This is further complicated by the fact that, under the Nice model, Jupiter and the other giant planets migrated significantly in their orbits during the earlier history of the \textcolor{red}{Solar System} \citep[e.g.,][]{morbidelli2005chaotic_Nice1, gomes2005origin_Nice2, tsiganis2005origin_Nice3}. 
Although the existence of pulsar Trojans is uncertain, the history of serendipitous astronomical discoveries in the last several decades---including pulsars, fast radio bursts, and hot Jupiters---motivates their search. In any case, our results provide an independent observational baseline for future theoretical studies, such as hydrodynamical simulations, of pulsar Trojan viability.

% Ever since the discovery of the first exoplanet \citep[][]{wolszczan1992_firstexoplanet}, there has been a growing interest in discovering massive bodies in the environs of pulsars. 
The first exoplanet was discovered around a pulsar \citep{wolszczan1992_firstexoplanet}, and ever since there has been a growing interest in discovering massive bodies in the environs of pulsars. 
Besides studies looking for planets in the ordinary P-type orbit \citep[e.g.,][]{bailes2011transformation, spiewak2018psr, behrens_nanograv_2020, nictu2022jodrellbankexoplanetsearch}, pulsar asteroid belts \citep[][]{shannon2013asteroidbelt, jennings_pulsar_2020_asteroid} and freely-floating planets gravitationally scattering with pulsars \citep[][]{jennings2020ISO, loki2025ISO} have also been investigated.
% We call these pulsar objects ZOMBIES (Zombies are Objects/Massive Bodies In the Environs of pulsarS).
% These works are part of what we call the ZOMBIES\footnote{\underline{Z}OMBIES are \underline{O}bjects/\underline{M}assive \underline{B}odies \underline{I}n the \underline{E}nvirons of pulsar\underline{S}} search.
In this work, we turn the search to Trojans that may reside in black widow pulsar systems---binaries characterized by an MSP with a low-mass ($\sim0.01\,\mathrm{M_\odot}$) companion \citep[e.g.,][]{shahbaz2017black_widow}---and are therefore likely to follow the stability criterion given in Equation~\ref{eq:gascheau_stability}.

To conduct the first-ever search for Trojan planets in pulsar binary systems, we use the North American Nanohertz Observatory for Gravitational Waves
(NANOGrav) 15-year (NG15yr) narrow-band dataset \citep[][]{agazie2023nanograv} as well as optical-band light curves of PSR~J1641+8049 from \citet{kirichenko_black_2024} to search for pulsar systems bearing Trojans in the T-type configuration. We ignore horseshoe configurations because these systems do not yield a closed-form waveform in the pulse times of arrival (TOAs; see Section~\ref{sec:libration_theory} and \citealt{leleu_mainpaper_2015}).
A Trojan in the T-type configuration librates around a particular triangular Lagrange point, while a Trojan in the horseshoe configuration will repeatedly librate between $L_4$ and $L_5$ going through $L_3$. Excluding horseshoe orbiters does not greatly limit our search, as not only are they less common in the Solar System \citep[e.g.,][]{kaplan2020horseshoe}, but also their stability criterion are stricter \citep[e.g.,][]{laughlin2002extrasolar}. 

We outline our methods in Section~\ref{sec:methods}, select our observed data in Section~\ref{sec:alldata}, display and discuss our results in Section~\ref{sec:results}, and conclude with a broader discussion in Section~\ref{sec:discussion}.

% All uncertainties reported in this work enclose the 68\% credible interval. We quote our upper limits at 95\% credibility. We use both the parentheses-on-the-last-digit(s) convention and the $\pm$ notation to express these 68\% uncertainties. Throughout, $\log$ denotes the base-10 logarithm.

\section{Methods}
\label{sec:methods}

\citet{leleu_mainpaper_2015, leleu_theory_alphamethod_2017} showed that the signal observed from a star-exoplanet-exotrojan system is largely degenerate with that from an ordinary star-exoplanet system. 
This is due to the secondary and Trojan forming a combined effective companion. The TOAs will indicate the pulsar and a combined effective companion are orbiting a center of mass.
The following subsections discuss two ways to break this degeneracy in the context of pulsar binary systems. 
% We ignore $\mathcal{O}(e_{\rm o})$ effects, where $e_o$ is the orbital eccentricity, throughout this study, because the highest orbital eccentricity of the pulsars studied is of the order $10^{-4}$.

The first method uses both optical light curves and radio pulsar TOAs. Both the light curves and TOAs can independently measure the time of superior conjunction (TSC), which is when the pulsar's companion is furthest from the Earth in its orbit. A discrepancy between the light-curve-measured TSC and TOA-measured TSC can reveal the presence of a mass accumulation along the orbit in the system. 
% We consider the cases of \textcolor{red}{$L_4$ or $L_5$} mass accumulations, for example, relatively diffuse gas or a swarm of several Trojans, and Trojans that are much denser than the secondary so that we can ignore the flux from a tidally-locked Trojan. A mass accumulation would not necessarily be tidally locked and therefore would not have a rigid day and night side. We will hereafter refer to a ``Trojan mass accumulation" as a Trojan as well for simplicity. 
This is an adaptation the method developed by \citet{ford_observational_2006} for exoplanets orbiting main-sequence stars.
In their case, they compare the transiting time of the secondary measured from photometry---in other words, the time of inferior conjunction (TIC)---with the estimated transiting time from radial velocities. The orbital planes of the secondaries in our study are not ``edge-on" with respect to our LOS, so we cannot measure their direct transits. However, we can measure their TSCs which corresponds to peak optical flux, as the secondaries are tidally locked with pulsar. Note that we assume negligible flux contributions from the pulsar and any Trojans. Analogously to radial velocities, TOAs trace the Keplerian orbit from the effective orbiting mass (secondary-exotrojan) by measuring the pulsar LOS position. Therefore, the TOAs uniquely determine the relevant Keplerian parameters \citep[e.g.,][]{APTBtaylor2024}, which give the TSC.
% In their case, they use photometric transits to measure the time of inferior conjunction (TIC) and compare it to the TIC predicted from radial-velocity curves. In our case, the secondary does not eclipse the pulsar, so we cannot measure the TIC, but the TSC is readily available. Additionally, we do not have radial velocity measurements from the pulsar, but rather TOAs which can directly measure the pulsar's position (relative to the system center of mass) along the Earth's line of sight (LOS) due to the changing light travel time. The TOAs can measure the relevant Keplerian parameters \citep[e.g.,][]{APTBtaylor2024}, which give the TSC.
See Section~\ref{sec:sup_conj} for more details and Figures~1A and 1C of \citealt{ford_observational_2006} for an illustration of the concept.

The second method requires only TOAs, but only works if any Trojan and secondary in the system is librating around the stable Lagrange point. 
% The TOAs will indicate the pulsar and a combined effective companion are orbiting a center of mass. 
If the Trojan and secondary are librating with some frequency, the effective companion-pulsar center of mass will also oscillate at the same frequency. A pulsar orbiting an oscillating center of mass no longer produces TOAs degenerate with those of a two-body system, and allows us to measure important properties and masses of the system. This method comes directly from \citet{leleu_mainpaper_2015}, where they derive both the line-of-sight position and radial velocity of a main-sequence star that hosts librating co-orbitals. As mentioned above, we can use the TOAs to measure the pulsar's line-of-sight (LOS) position. We expand upon their LOS position derivation in Appendix~\ref{app:libration_derivation}, including a minor algebraic correction. 
This libration-dependent method can be used to detect a single Trojan in a pulsar binary system. In the case of a Trojan swarm, this method remains effective in detecting the highest-mass Trojan, provided this Trojan is massive enough to allow gravitational effects from the other Trojans to be ignored. 
See Section~\ref{sec:libration_theory} for more details. 

In this work, we quote posterior medians with uncertainties enclosing the 68\% credible interval. We quote our upper limits at 95\% credibility. We use both the parentheses-on-the-last-digit(s) convention and the $\pm$ notation to express these 68\% uncertainties. Throughout, $\log$ denotes the base-10 logarithm.

% In some black widow systems, the pulsar is heating its companion up to several thousand K \citep[e.g.,][]{kirichenko_black_2024}. Due to the companion being tidally locked to the pulsar, the companion has strict heated and non-heated sides, which we will call the day and night sides, respectively. When the companion is furthest from the Earth, known as superior conjunction, its day side will face the Earth (up to the sine of the orbital inclination). Therefore, we can measure the time of superior conjunction (TSC) by measuring the time of peak optical flux from the system. The pulse TOAs can independently measure the TSC. A discrepency between the TOA-measured TSC and the optical-flux-measured TSC can be attributed to a Trojan. 

\subsection{Conjunction Inconsistency}
\label{sec:sup_conj}

% The reflex motion of the pulsar due to the orbit of its companions is fully consistent with the replacement of the secondary and exotrojan with a single effective object \citep{leleu_theory_alphamethod_2017}. If the time of the secondary's furthest (closest) approach to Earth---known as superior (inferior) conjunction---can be measured directly through optical band data, then this time, $T_{\rm o,s}$ ($T_{\rm o,i}$), can be compared to the conjunction time as predicted by the pulsar's reflex motion, $T_{\rm r,s}$ ($T_{\rm r,i}$). 
Let $T_{\rm o,s}$ and $T_{\rm o,i}$ be the times of superior and inferior conjunction, respectively, as measured through optical band data, and let $T_{\rm r,s}$ and $T_{\rm r,i}$ be the same as measured by the TOA-determined pulsar reflex motion.
% As described above, we are inspecting any difference between . 
% See Figures~1A and 1C of \citealt{ford_observational_2006} for an illustration of the concept. 
Given how \citet{lange2001precision} define $T_{\rm asc}$, radio pulsar timing predicts superior and inferior conjunctions to occur at $T_{\rm r,s} = T_{\rm asc} - P_{\rm b}/4$ and $T_{\rm r,i} = T_{\rm asc} + P_{\rm b}/4$, respectively, where we ignore orbital eccentricity ($e_{\rm o}$) effects. $P_{\rm b}$ is the binary orbital period and $T_{\rm asc}$ is the time of the ascending node as defined in \citet{lange2001precision}.
We ignore $\mathcal{O}(e_{\rm o})$ effects throughout this study, because the highest orbital eccentricity of the pulsars studied is of the order $10^{-4}$.
\citet{ford_observational_2006} show that any discrepancy in $T_{\rm o,s}$ and $T_{\rm r,s}$ (or $T_{\rm o,i}$ and $T_{\rm r,i}$) could indicate the presence of an exotrojan with the time difference directly mapped from the exotrojan mass: 
% \begin{equation}
%     \label{eq:sup_conj}
%     \Delta t\approx 0.11 \left( \frac{P}{3 \:\text{hours}} \right) \left( \frac{0.04 M_\odot}{m_T + m_c} \right) \left( \frac{m_T}{1\:M_\oplus} \right)\: \text{seconds}
% \end{equation}
\begin{equation}
    \label{eq:sup_conj}
    \Delta t\approx \pm 35.6 \left( \frac{P_{\rm b}}{3 \,\text{h}} \right) \left( \frac{0.04 \, {\rm M}_\odot}{m_{\rm T} + m_{\rm c}} \right) \left( \frac{\Delta m_{\rm T}}{1\,\rm{M}_{\rm J}} \right)\, \text{s},
\end{equation}
where $\Delta t = T_{\rm o,s} - T_{\rm r,s}$ and $\Delta m_{\rm T}$ is the mass difference between the $L_4$ and $L_5$ Lagrange points. Here, again, we ignore the $\mathcal{O}(e_{\rm o})$ term. Therefore, Trojans present in both points will partially cancel their measurable effects. We use this method to rule out a high-mass Trojan in the PSR~J1641+8049 system (see Section~\ref{sec:results_conjunction}) by incorporating its superior conjunction measurements (Section~\ref{sec:optical_data}). We note that this method is inclination independent due to the orbital period and the time delay both being inclination independent.

% \begin{figure}
%         \centering
% 	\includegraphics[width=\columnwidth]{conjunction.pdf}
% 	% \includegraphics[width=\textwidth]{conjunction.pdf}
%     \caption{A schematic of the positions of the Trojan, $m_{\rm T}$, the secondary, $m_{\rm c}$, and the effective combined object, $m_{\rm f}$. $\Delta \phi$ is the relative angle between the orbital longitudes of the combined companion and the actual secondary. The Trojan and secondary are shown orbiting counterclockwise here, meaning the Trojan is in the $L_4$ Lagrange point of the pulsar-secondary system. At the time of this figure, the companion is at its closest position to Earth, known as inferior conjunction. The line of sight to the Earth is shown in red. See the text for more details. Inspired by Figure 1 of \citet{ford_observational_2006}.}
%     \label{fig:conjunction}
% \end{figure}

\subsubsection{Determination of the Times of Optical Superior Conjunction}

Some binary pulsar optical light curves show a clearly periodic structure with a period equal to the binary period (e.g., Figure~4 of \citealt{kirichenko_black_2024}), explained by the neutron star heating its tidally locked companion. Therefore, assuming the symmetric heating models of \citet[][]{zharikov2013accretion, zharikov2019optical} and neglecting the heating of the Trojan, the peak flux corresponds to the time(s) of superior conjunction of the secondary.

\begin{figure}
        % \centering
	\includegraphics[width=\columnwidth, bb = 0 0 225 157, clip=]{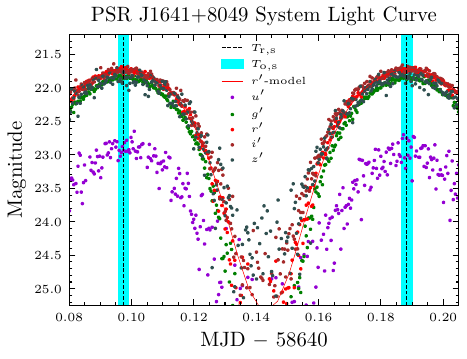}
    \caption{High PERformance CAMera light curve data from the 10.4m Gran Telescopio Canarias of the PSR~J1641+8049 system for each Sloan band. The data was first presented in \citet{mata2023black0023}. We show the best-fit model for the $r'$ band as an example, with the remaining model curves shown in \citet{kirichenko_black_2024}. Every band except $u'$ was used in calculating $T_{\rm o, s}$. The vertical dashed line marks the TSC as predicted by radio timing. The uncertainty on the TOA-predicted TSC is far smaller than the $\hat{x}$-axis scale of the figure. The vertical cyan band encloses the 68\% confidence interval of the TSC as calculated from the optical light curve.}
    \label{fig:optical_light_curve}
\end{figure}

To derive the times of the superior conjunction and their associated uncertainties, we analyzed light curve data of the PSR~J1641+8049 system. We show the optical light curve in Figure~\ref{fig:optical_light_curve}.
More details can be found in Section~\ref{sec:optical_data} and \citet{mata2023black0023, kirichenko_black_2024}.
We adopt the emission model of \citet{zharikov2013accretion, zharikov2019optical} and start with the best-fit parameters found in \citet{kirichenko_black_2024}. The emission model assumes symmetric heating with hottest side facing the pulsar. 
The heating model parameters are the distance from the system to the Earth, the reddening $E(B-V)$, the pulsar mass, the companion-to-pulsar mass ratio, the orbital inclination, the Roche lobe filling factor, the effective radiation factor, the companion dark-side temperature, and the time of superior conjunction. For more details on these parameters and the model, see \citet{zharikov2013accretion, zharikov2019optical, kirichenko_black_2024} and references therein. We define $\mathbf{p}$ as the vector of all these parameters, excluding the time of superior conjunction.

Let $\phi_{\rm max}$ be the binary phase corresponding to peak light, which is used to calculate the time of superior conjunction. Let $\phi_{\rm max}'$ and $\mathbf{p}'$ be the values of $\phi_{\rm max}$ and $\mathbf{p}$, respectively, that minimize $\chi^2$. While fixing $\mathbf{p}$ to $\mathbf{p}'$, we shift $\phi_{\rm max}$ from $\phi_{\rm max}'$ and compute the $\Delta \chi^2$ for each shift, $\Delta \chi^2 (\phi_{\rm max})$. 
The resulting $\Delta \chi^2$ values form a distribution, $f(\Delta \chi^2)$. The 68\% uncertainty on $\phi_{\rm max}$, $\sigma_{\phi_{\rm max}}$, is such that
\begin{equation}
    \int_0 ^{\Delta \chi^2(\sigma_{\phi_{\rm max}})} f(\Delta \chi^2) \, \rm{d}\Delta \chi^2 = 0.68.
\end{equation}

% The resulting $\Delta \chi^2$ values follow a $\chi^2$ distribution \citep{wilks1938large}. The 68\% confidence interval on this distribution corresponds to the $\phi_{\rm max}$ shift that gives that interval.

% Using Wilks' theorem \citep{wilks1938large}, we report the shift that yields $\Delta \chi^2 = 1$ as the 68\% uncertainty interval (see, e.g., \citealt{modern_wilks_usage2025comparison} for a more recent use of Wilks' theorem).

% from the best fit $\mathbf{p}$, which we will call $\mathbf{p}'$. While fixing $\mathbf{p}$ to $\mathbf{p}'$, we shift 

% While fixing $\vec{p}$ on that of the best-fit model, we shift the zero binary phase from the zero phase in the best-fit model and compute the $\Delta \chi^2$ for each shift. Using Wilks' theorem \citep{wilks1938large}, we report the shift that yields $\Delta \chi^2 = 1$ as the 68\% uncertainty interval (see, e.g., \citealt{modern_wilks_usage2025comparison} for a more recent use of Wilks' theorem). 

\subsection{Librations}
\label{sec:libration_theory}

\citet{leleu_mainpaper_2015} adapted the work of \citet{robutel2011analytical} to determine how a host star's position (or velocity if spectral data is available) relative to the Solar System barycenter (SSB) will change when a companion and Trojan are present. A pulsar moving relative to the SSB will cause changing light travel time delays, known as R{\o}mer delays, in the pulse TOAs. 
\citet{jennings2020ISO}, for example, showed these delays to be
\begin{equation}
\label{eq:ross_toa_delay}
    \Delta \tau = -\frac{\boldsymbol{r}_{\rm p} \cdot \boldsymbol{\hat{n}}}{c}
\end{equation}
when ignoring terms that are degenerate with the pulsar's pulse period and pulse period time derivative. $\boldsymbol{r}_{\rm p}$ is the pulsar's position with respect to the system center of mass, $\boldsymbol{\hat{n}}$ is the unit vector pointing from the pulsar system barycenter to the SSB, and $c$ is the speed of light in a vacuum. Therefore, $\boldsymbol{\hat{n}}$ is antiparallel to the LOS. 

We will now begin the derivation of $\Delta \tau$ in the case of a librating Trojan in the orbit around a pulsar, but reserve most of the details to Appendix~\ref{app:libration_derivation}. Let $\lambda_{\rm T}$ and $\lambda_{\rm c}$ be the orbital longitudes of the Trojan and secondary, respectively. Let $\lambda$ be the mass-averaged longitude of the co-orbitals which follows the mean motion and is unaffected by librations.
We call the difference in longitude between the co-orbitals the libration angle, $\zeta$. The TOAs will indicate the presence of an orbiting combined effective companion, which has longitude $\phi$. $\lambda$ and $\phi$ are nearly but not exactly identical, with 
\begin{equation}
|\phi - \lambda| \lesssim 0.02
\end{equation}
shown in Appendix~\ref{app:libration_derivation}.
Both $\lambda$ and $\phi$ evolve over time as the co-orbitals orbit the pulsar according to
\begin{align}
    \lambda = n_{\rm b} \tilde{t} + \lambda_0, \\
    \phi = n_{\rm b} \tilde{t} + \phi_0, 
\end{align}
where $n_{\rm b} = 2\pi/P_{\rm b}$ is the orbital mean motion, $\tilde{t}$ is time with respect to the time of the ascending node, $T_{\rm asc}$, and $\lambda_0$ and $\phi_0$ are to-be-determined constants.
We show in Appendix~\ref{app:libration_derivation} that the mass of the fictitious effective companion
\begin{equation}
\label{eq:mf}
m_{f} =(m_{\rm c} + m_{\rm T})\sqrt{1 - \delta (1- \delta)} + \mathcal{O}\left(\frac{m_{\rm T}}{m_{\rm p}},\,\frac{m_{\rm c}}{m_{\rm p}}\right).
\end{equation}
Defining 
\begin{equation}
    \delta = m_{\rm c}/(m_{\rm c} + m_{\rm T}),
\end{equation}
the mass of the secondary relative to both co-orbitals, the angles of the Trojan and secondary longitudes with respect to $\lambda$ are $\delta \zeta$ and $(1-\delta) \zeta$, respectively.
\citet{robutel2011analytical} showed that
\begin{align}
    \lambda_{\rm T} &= \lambda + \delta \zeta, \\
    \lambda_{\rm c} &= \lambda - (1-\delta) \zeta.
\end{align}

\begin{figure}
        % \centering
	\includegraphics[width=\columnwidth]{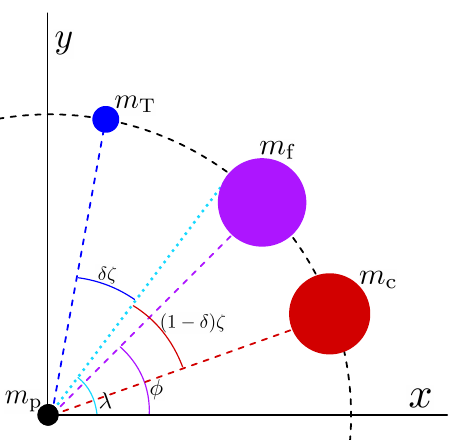}
    \caption{The relevant angles in the triangular three-body system. $m_{\rm f}$ indicates the fictitious companion, with mass given by Equation~\ref{eq:mf}. The Trojan and companion are shown orbiting counterclockwise here, meaning the Trojan is in the $L_4$ Lagrange point. See the text for more details. Inspired by Figure 1 of \citet{leleu_mainpaper_2015}.}
    \label{fig:libration_diagram}
\end{figure}

Let the plane of the orbit define the $\hat{x}\hat{y}$-plane. Without a loss of generality, we place the Earth in the $\hat{x}\hat{z}$-plane and define the $\hat{x}$-axis to be the projection of the LOS onto the orbital plane. Therefore, the orbital inclination with respect to the plane of the sky, $I$, is the angle between the LOS and the $\hat{z}$-axis. We define the $\hat{y}$-axis such that the $T_{\rm asc}$ is the time at which the fictitious companion crosses the negative $\hat{y}$-axis. See Figure~\ref{fig:libration_diagram} for a helpful illustration. 

When librations are absent, $\zeta = \pi/3$ for all time. However, in the presence of small librations, $\zeta$ will oscillate around $\pi/3$ with (angular) frequency $\nu$ and amplitude $z$ \citep[e.g.,][]{leleu_mainpaper_2015}; i.e.
\begin{equation}
    \label{eq:zeta_approx}
    \zeta = z\cos{(\nu \tilde{t} + \varphi)} + \frac{\pi}{3},
\end{equation}
where $\varphi$ is the vibrational phase at $t = T_{\rm asc}$.
$z$ can vary across otherwise identical systems and is a function of the Trojan's detailed formation history. We note that $z$ can be lowered through any sort of dissipative process, such as gas drag \citep{lyra2009_trojanearthformation}. In contrast, $\nu$ depends on the masses of the pulsar and co-orbitals. Namely, \citet{leleu_mainpaper_2015} showed that in the low $\mu$ limit,
\begin{equation}
    \label{eq:nu_leleu}
    \nu = n_{\rm b} \sqrt{\frac{27 \mu}{4}},
\end{equation}
where
\begin{equation}
    \label{eq:mu_definition}
    \mu = \frac{m_{\rm T}+m_{\rm c}}{m_{\rm p}+m_{\rm c} + m_{\rm T}}.
\end{equation}
However, in the low-$m_{\rm T}$-only limit 
\begin{equation}
    \label{eq:nu_cornish}
    \nu = \frac{n_{\rm b}}{2} \left[ 2 - \sqrt{ 27\kappa^2 - 23 } \right]^{1/2},
\end{equation}
where $\kappa = (m_{\rm p}- m_{\rm c})/(m_{\rm p} + m_{\rm c})$ \citep[][]{cornish_lagrange, erdi2007secondary}. Equation~\ref{eq:nu_cornish} is more appropriate than Equation~\ref{eq:nu_leleu} when $m_{\rm T}$ is negligible but $m_{\rm c}$ is not. There is another frequency associated with librations as well \citep[][]{erdi2007secondary}, which would add in sum to Equation~\ref{eq:zeta_approx}, but \citet{janson_kepler_indivdual2013} showed that its corresponding libration amplitude is much smaller than $z$, and therefore can be safely ignored when these systems are on the threshold of detectability. 

\begin{figure}
        % \centering
	\includegraphics[width=\columnwidth]{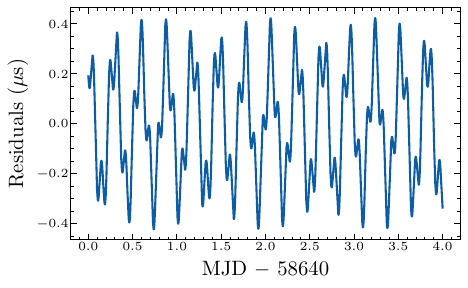}
    \caption{Model for the effect of a librating Trojan on the TOA delays (residuals) of PSR~J0023+0923. In this example, we set $m_{\rm T} = 6$\,M$_{\rm J}$, $m_{\rm p} = 1.4$\,M$_{\odot}$, $m_{\rm c} = 46$\,M$_{\rm J}$, $I=14^\circ$, $z=0.0001^\circ$, $\varphi = 1.2$. The other relevant parameters are either determined by already-known binary system parameters of PSR~J0023+0923 (Table~\ref{tab:pulsar_parameters}) or the equations in Section~\ref{sec:libration_theory}. Besides the Table~\ref{tab:pulsar_parameters} parameter values, these are $\mathcal{T} \approx 0.42\,\mu$s, $\nu \approx 24\,$rad/d, and $\theta \approx 1.147$. If the PSR~J0023+0923 system had a Trojan with these properties, the TOA residuals would be of the order of tenths of $\mu$s, consistent with $\mathcal{T}$.}
    \label{fig:libration_example}
\end{figure}

If $z=0$, then the TOAs are completely degenerate from TOAs from a two-body system. A non-zero $z$ breaks this degeneracy. 
The TOA signal, through Equation~\ref{eq:ross_toa_delay}, will have a term degenerate with a two-body system plus a term due to librations. 
% By changing the relative position of the pulsar, the TOA signal, through Equation~\ref{eq:ross_toa_delay}, 
This ``libration term" will depend on the mass-averaged semi-major axis of the co-orbitals in the pulsar-centric frame ($a$), $z$, $\nu$, $n_{\rm b}$, $I$, and $\mu$. After removing the degenerate two-body term, we derive in Appendix~\ref{app:libration_derivation} 
% \begin{equation}
%     \label{eq:libration_residuals_full}
%     \Delta \tau=\frac{\mu a \sin{I}}{c} z \delta (1-\delta)\left[ \cos{(\nu \tilde{t} + \varphi)} \cos{(n_{\rm b}\tilde{t} + \theta)} + \frac{2}{3}\frac{\nu}{n_{\rm b}} \sin{(\nu \tilde{t} + \varphi)} \sin{(n_{\rm b}\tilde{t} + \theta)} \right],
% \end{equation}
\begin{align}
\label{eq:libration_residuals_full}
% \begin{split}
\Delta \tau =& \frac{\mu a \sin{I}}{c} z \delta (1-\delta)\Big[\cos{(\nu \tilde{t} + \varphi)} \cos{(n_{\rm b}\tilde{t} + \theta)} \nonumber \\
&+ \frac{2}{3}\frac{\nu}{n_{\rm b}} \sin{(\nu \tilde{t} + \varphi)} \sin{(n_{\rm b}\tilde{t} + \theta)}
\Big],
% \end{split}
\end{align}
where 
\begin{equation}
    \label{eq:theta}
    \theta = \arctan\left(\frac{\delta \sqrt{3}}{2 - \delta}\right) + \frac{\pi}{3}. 
\end{equation}
In searching for a Trojan in a particular system, see Section~\ref{sec:bayesian_parameters}, we wish to remove covariances and degeneracies in the prefactor of Equation~\ref{eq:libration_residuals_full}, so we define what we call the libration strength
\begin{equation}
    \label{eq:T}
    \mathcal{T} = \alpha z \delta (1 - \delta),
\end{equation}
where
\begin{equation}
    \alpha = \frac{\mu a}{c} \sin{(I)}.
\end{equation}
Therefore, we have
\begin{align}
    \label{eq:libration_residuals}
    \Delta \tau=&\mathcal{T}\Big[ \cos{(\nu \tilde{t} + \varphi)} \cos{(n_{\rm b}\tilde{t} + \theta)} \nonumber \\
    &+ \frac{2}{3}\frac{\nu}{n_{\rm b}} \sin{(\nu \tilde{t} + \varphi)} \sin{(n_{\rm b}\tilde{t} + \theta)} \Big].
\end{align}
We plot this function for an example set of parameters in Figure~\ref{fig:libration_example}. The timing model, Equation~\ref{eq:ross_toa_delay}, would cause TOA residuals, which are the measured TOAs minus the predicted TOAs, if a Trojan in the system was not accounted for. 

While pulsar timing cannot reveal the semi-major axis of the pulsar's orbit around the system center of mass ($a_{\rm p}$) directly, it can measure its projection onto the plane of the sky 
\begin{equation}
K \equiv \frac{a_{\rm p}}{c} \sin{I}.
\end{equation}
This is useful quantity because it can be related to the masses of the co-orbitals through 
\begin{equation}
    \label{eq:K}
    K = \alpha \sqrt{1 - \delta (1 - \delta)}
\end{equation}
by setting $K$ to be the amplitude of the two-body degenerate term of the three-body system (see Appendix~\ref{app:libration_derivation}). The value of $K$ is known quantity for any timed binary pulsar \textit{a priori} to any Trojan search. 

In a system with an already known $n_{\rm b}$ and $K$, measuring the four libration parameters---$\mathcal{T}$, $\nu$, $\theta$, and $\varphi$, the additional parameters required to describe Equation~\ref{eq:libration_residuals}---can reveal many physical properties of the system.
To do so, we solve Equations~\ref{eq:theta}, \ref{eq:K}, and \ref{eq:T} analytically for $\delta$, $\alpha$, and $z$:
\begin{align}
\delta &= \frac{2\beta}{\sqrt{3} + \beta} \text{ where } \beta = \tan{(\theta - \pi/3)},\label{eq:delta} \\
\alpha &= \frac{K}{\sqrt{1 - \delta(1-\delta)}}, \label{eq:alpha}\\
z &= \frac{\mathcal{T}}{\alpha \delta(1-\delta)}. \label{eq:z}
\end{align}
Therefore, Equation~\ref{eq:delta} shows that $\theta$ gives a direct measure of the relative Trojan mass, $\delta$.
Upon assuming a pulsar mass, the companion and Trojan masses are 
\begin{equation}
    \label{eq:m_c_m_p_kappa}
    m_{\rm c} = \frac{1-\kappa}{1+\kappa} m_{\rm p},
\end{equation}
\begin{equation}
    m_{\rm T} = \frac{\delta}{1-\delta}m_{\rm c}.
\end{equation}
The orbital inclination can be determined too, using the Newtonian binary mass function
\begin{equation}
    \label{eq:binary_mass_function}
    \frac{f}{\rm{M}_\odot}= \frac{(m_{\rm c} \sin{I})^3}{(m_{\rm p} + m_{\rm c})^2} = \frac{K^3}{T_\odot}n_{\rm b}^2,
\end{equation}
where $T_\odot = 4.925490947641 ...\, \rm{\mu s}$ is the nominal solar mass parameter.\footnote{This is a defined quantity. See Section~3.4.1 of \citet{freire_wex2024gravity_binary} for a brief discussion on $T_\odot$.}

\subsubsection{Determination of Libration Parameters}
\label{sec:bayesian_parameters}
We adapt the search methodology of \citet{loki2025ISO}, originally developed for freely floating interstellar objects (FFOs), to our Trojan search. In order to seek out the libration delays of Equation~\ref{eq:libration_residuals}, we model not only $\Delta \tau$, but also white noise, red noise, and the ordinary pulsar timing model, which includes parameters such as pulsar spin period, spin period derivatives, sky position, and binary parameters (see \citealt{edwards2006tempo2, luo2019ascl, luo2021pint, susobhanan2024pint} for more-detailed discussions on the ordinary pulsar timing model). We model the noise and timing model in the exact same way as \citet{loki2025ISO}.
Noise and timing modeling are handled in the \texttt{ENTERPRISE} Python library \citep{EllisVallisneri+2017,JohnsonMeyers+2023}. We also use the \texttt{PTMCMCSampler} Python library \citep{EllisvanHaasteren2017,JohnsonMeyers+2023}, which is an implementation of the parallel tempering Markov-Chain Monte Carlo (PTMCMC) method (see, for example, \citealt[][]{earl2005parallel, vousden2016dynamic} for discussions on PTMCMC). Unlike \citet{loki2025ISO}, we used sixteen temperature chains rather than just one because the posterior space in our case can be highly multimodal (see Section~\ref{sec:J0023_librations}).

We do not have to worry about degeneracies between the noise modeling and the libration frequency. The red noise is modeled as frequencies in integer multiples of $1/T_{\rm span}$, where $T_{\rm span}$ is the time span of TOAs for a given pulsar, up to $30/T_{\rm span}$. The highest such frequency for the eight NG15yr pulsars of interest is $30 / (3.4\,{\rm yr})\sim 9\,\rm{yr}^{-1}$. The libration frequencies are of order $1\,\rm{day}^{-1}\sim 400 \, \rm{yr}^{-1}$ (see Figures~\ref{fig:J0610_post}, \ref{fig:J0023_post}, and \ref{fig:J1705_post}). Therefore, the red noise frequency space and the libration frequency space are disjoint. 

\subsubsection{Libration Priors}
\label{sec:priors}
We adopt uninformative priors on the libration parameters, $\mathcal{T}$, $\nu$, $\theta$, and $\varphi$. We also adopt the same priors on the red noise as in \citep[][]{loki2025ISO}. We do not search for $n_{\rm b}$ in our MCMC approach as the mean motion is already well constrained by the timing model.

\begin{itemize}
    \item We choose a log-uniform distribution for $\mathcal{T}$ of $\log{(\mathcal{T}/\rm{s})}\sim \mathcal{U}(-12, \,-4)$. The lower limit of $-12$ was chosen because such a signal strength is far beneath the noise floor of NG15yr MSPs, which is on the order of 0.1--1 microseconds \citep[][]{agazie2023nanograv}. The $-4$ upper limit is in place because a signal of such strength would already easily be detected by eye in the TOA residuals. 

\item Our prior on $\nu$ is $\nu \sim \mathcal{U}(\nu _{\rm min}, \,\nu _{\rm max})$. The maximum real-valued $\nu$ in Equation~\ref{eq:nu_cornish} is $n_{\rm b} / \sqrt2$. Therefore, this is our $\nu _{\rm max}$. The minimum value, $\nu _{\rm min}$, is determined by finding the smallest $\mu$ for a given binary mass function (Equation~\ref{eq:binary_mass_function}) and inserting that $\mu$ into Equation~\ref{eq:nu_leleu}. Such a $\mu$ is found by setting $\sin{I} = 1$ and the pulsar mass to the most massive discovered pulsar ($\sim2.5\,\mathrm{M}_\odot$; \citealt{Clifford2019BS}).

\item We choose the prior $\theta \sim \mathcal{U}(\pi/3, \,2\pi/3)$ because these bound the physical range of $\theta$ according to Equation~\ref{eq:theta}, given $\delta \in [0,\, 1]$. We also investigate the $\theta \sim \mathcal{U}(0, \,2\pi)$ prior to account for any nonphysical signals that may enter our physically-motivated prior space due to uncertainties in $\theta$. 
Lastly, we look in $\log{\xi}$ space, where $\theta = \pi/3 + \xi$, with a $\log{\xi} \sim \mathcal{U}(-6.2,\,\log{(\pi/3)})$ prior. The minimum of $-6.2$ is chosen because this gives the corresponding $\theta$ when the Trojan is 1 lunar mass and the secondary is $0.05\,\rm{M}_\odot$, the most massive companion satisfying Equation~\ref{eq:gascheau_stability} when $m_{\rm p} = 1.35\,\rm{M}_\odot$. A search in $\log{\xi}$, where 
\begin{equation}
    \xi = \theta - \pi/3,
\end{equation}
allows us to probe small Trojan masses in the $L_5$ Lagrange point (Section~\ref{sec:J0023_librations}). We do not carry out an analogous $\theta = 2\pi/3 - \xi'$ search over a uniform $\log{\xi'}$ prior because none of our positive detections are near $\theta = 2\pi/3$.
% I also look in log(eta) space for theta = \pi/3 + eta where log(eta) goes from -8 to log(pi/3)

\item We choose the prior $\varphi \sim \mathcal{U}(0, \,2\pi)$ because $\varphi$ is an arbitrary phase. We take special care to ensure $\varphi$ is $2\pi$-periodic in our use of \texttt{PTMCMCSampler}.
\end{itemize}

\section{Selected Pulsars and their Observational Data}
\label{sec:alldata}

According to the Australia Telescope National Facility Pulsar Catalog \citep[][]{manchester2005australia},\footnote{\url{https://www.atnf.csiro.au/research/pulsar/psrcat}} there are 72 pulsar binary systems that satisfy Equation~\ref{eq:gascheau_stability} when $m_{\rm T}=0$, $m_{\rm p} = 1.35\,$M$_\odot$, and $m_{\rm c}$ is determined by the Newtonian mass function (Equation~\ref{eq:binary_mass_function}) when $\sin{I} = 1$, where $I$ is the pulsar system's orbital inclination with respect to the plane of the sky. 
We define $m_{\rm c,\, \min}$ as the minimum secondary mass when $m_{\rm p} = 1.35\,$M$_\odot$. Note that $m_{\rm c,\, \min} = m_{\rm c}$ when $I = 90^\circ$.
Choosing $m_{\rm T}=0$ in evaluating Equation~\ref{eq:gascheau_stability} gives us systems that may have Trojans, provided they are not too massive. Any discovered pulsar Trojans would have non-zero masses that still satisfy Equation~\ref{eq:gascheau_stability}. Out of these 72 systems, we restrict this study to PSR~J1641+8049 and the eight systems that are in the NG15yr dataset (see Table~\ref{tab:pulsar_parameters}) due to their high observation cadence, long timing baseline, and availability of the dataset. Other datasets also satisfy these criteria \citep[e.g.,][]{antoniadis2023epta_second, zic2023parkes_ppta}, but we leave those analyses for future work due to differences in the noise modeling and observatories.
Useful properties of these pulsars can be found in Table~\ref{tab:pulsar_parameters}.
For PSR~J1641+8049, we use the method developed by \citet[][]{ford_observational_2006} to seek out Trojan-caused inconsistencies between the pulsar-companion conjunction times as measured by optical lightcurves of the system \citep[][]{kirichenko_black_2024} and those predicted by radio pulsar timing (see Section~\ref{sec:sup_conj}). For the eight NG15yr pulsars of interest, we adapt the method of \citet[][]{leleu_mainpaper_2015} to look for TOA residuals caused by Trojan librations---a method that only requires TOAs (see Section~\ref{sec:libration_theory}).

\begin{table*}
\centering
\caption{Relevant parameters of the binary pulsars in this study. Parentheses give 1$\sigma$ uncertainties on the trailing digit. The RMS is the root-mean-squared of the pulse TOA residuals. If $m_{\rm p} = 1.35 \,\rm{M}_\odot$, then the system can only contain Trojans if $m_{\rm c}$ is no higher than $0.05\,{\rm M}_\odot$, according to Equation~\ref{eq:gascheau_stability}. The eight pulsars after PSR~J1641+8049 are in the NG15yr dataset. We used TOAs from the Gran Telescopio Canarias (GTC), Canadian Hydrogen Intensity Mapping Experiment (CHIME), Arecibo Observatory (AO), and Green Bank Telescope (GBT).}
\begin{tabular}{lccccccc}
\hline
PSR & $P_{\rm b}$ (hr) & $K$ (lts) & $e_o$ & $T_{\rm span}$ (yr) & RMS ($\mu$s) & $m_{{\rm c},\,\min}$ (${\rm M}_\odot$) & Telescope \\
\hline
J1641+8049 & 2.1809748921(3)  & 0.0640753(2) & $2.2(4) \times 10^{-5}$ & 2.8 & 1.88 & 0.0404 & GTC, CHIME \\

J0023+0923  & 3.3311794144(4) & 0.03484142(4)
            & $7(2) \times 10^{-6}$
            & 9.0 & 6.02 & 0.0164 & AO \\

J0610$-$2100    & 6.864384046(3)  & 0.0734890(2)
                & $1.8(8) \times 10^{-5}$
                & 3.4 & 9.41 & 0.0214 & GBT \\

J0636+5128  & 1.5972322096(4) & 0.00898636(5)
            & $1.7(7) \times 10^{-5}$
            & 6.3 & 6.04 & 0.0069 & GBT \\

J1705$-$1903    & 4.414896263(2)  & 0.1043598(1)
                & $3(1) \times 10^{-6}$
                & 3.7 & 3.07 & 0.0412 & GBT \\

J1719$-$1438    & 2.17695076(2)   & 0.0018216(4)
                & $5(2) \times 10^{-4}$
                & 3.4 & 15.80 & 0.0011 & GBT \\

J1745+1017      & 17.52579411(2)  & 0.0881697(6)
                & $2(1) \times 10^{-5}$
                & 4.5 & 14.46 & 0.0137 & AO \\

J2214+3000      & 9.999190702(3)  & 0.0590815(2)
                & $2.1(6) \times 10^{-5}$
                & 8.4 & 7.31 & 0.0133 & AO \\

J2234+0944      & 10.071840895(2) & 0.0684298(1)
                & $7(3) \times 10^{-6}$
                & 7.1 & 5.76 & 0.0154 & AO \\
\hline
\end{tabular}
\label{tab:pulsar_parameters}
\end{table*}

\subsection{Optical Data}
\label{sec:optical_data}

The first optical observations of the PSR~J1641+8049 counterpart were reported by \citet{lynch2018}, revealing strong brightness variations tied to the binary period. The detection triggered further optical studies: using the High PERformance CAMera (HiPERCAM) installed at the 10.4m Gran Telescopio Canarias (GTC), \citet{mata2023black0023} performed phase-resolved multi-band photometric observations and analysed the light curves of the system, yielding its binary parameters and the distance.  %during two observing runs in the Sloan g$^\prime$-, r$^\prime$-, and i$^\prime$- bands with the Optical System for Imaging and low Resolution Integrated Spectroscopy (OSIRIS) instrument at the GTC.
\citet{kirichenko_black_2024} carried out independent phase-resolved multi-band optical studies with the Optical System for Imaging and low Resolution Integrated Spectroscopy (OSIRIS) at the GTC and performed a comparative light-curve analysis using both the HiPERCAM and OSIRIS datasets. %The details of the observations and data analyses were reported by \citet{2024MNRAS.527.4563K}. 

As demonstrated by both studies, %the shapes of the 
the PSR~J1641+8049 system light curves show a single-peaked sinusoidal-like modulation per orbital period with a maximum when the side of the companion heated by the neutron star is directed toward the observer (superior conjunction) and a minimum when the cold back side is visible (inferior conjunction). The system parameters obtained by \citet{kirichenko_black_2024} through the light curve analysis, including the distance to the source 4.6--4.8\,kpc and the orbital inclination 56--59$^\circ$, are found to be consistent with those of \citet{mata2023black0023}. 
However, the optical flux of the source at the maximum brightness phase faded by a factor of $\sim 2$ as compared with the values reported in the \citet{mata2023black0023} study.
Despite this variability, the companion’s irradiated hemisphere remains among the hottest (8000--9500\,K) observed in the known black widow systems.  

\subsection{Radio Pulsar Timing Data}
\label{sec:radio_pulsar_timing_data}

The times of arrival for PSR~J1641+8049 came from the CHIME/Pulsar backend \citep{amiri2021chime_CHIMEpulsar} on the Canadian Hydrogen Intensity Mapping Experiment (CHIME) radio telescope \citep{amiri2022overview_CHIME_tele}. These were used in calculating the times of superior conjunction.

\begin{figure}
        % \centering
	\includegraphics[width=0.8\columnwidth]{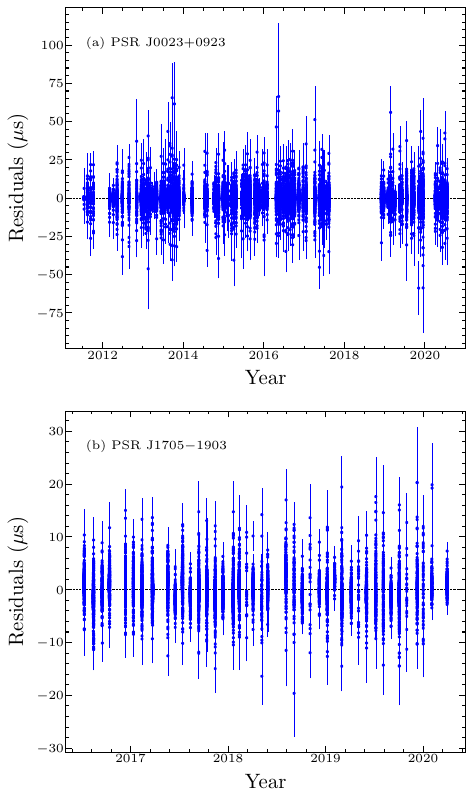}
    \caption{The pulse TOA residuals for PSRs~J0023+0923 and J1705$-$1903. The error bars indicate the 68\% uncertainty. These are the residuals after the ordinary timing model, but not red noise, has been accounted for (Section~\ref{sec:bayesian_parameters}). These two pulsars stand out above the others because their TOAs contain minor evidence for the Trojan libration signal (see Section~\ref{sec:results_librations}).}
    \label{fig:residuals_real}
\end{figure}

Of the eight selected NG15yr pulsars, four were observed with the Arecibo Observatory (AO; PSRs~J0023+0923, J1745+1017, J2214+3000, and J2234+0944) and four with the Green Bank Telescope (GBT; PSRs~J0610$-$2100, J0636+5128, J1705$-$1903, and J1719$-$1438).
The TOAs of our selected pulsars cover time spans ($T_{\rm span}$) ranging from 3.4\,yr (PSR~J0610$-$2100) to 9.0\,yr (PSR~J0023+0923). The near-random telescope cadences with respect to libration phase (see Equations~\ref{eq:zeta_approx} and \ref{eq:libration_residuals}), combined with the years-long timing baselines, ensure that the TOAs effectively sample all libration phases. Therefore, the Nyquist–Shannon sampling theorem criterion is relaxed, and the libration frequencies that are physically allowed (Section~\ref{sec:priors}) can be well-constrained. We show the TOA residual time series for PSRs~J0023+0923 and J1705$-$1903 in Figure~\ref{fig:residuals_real} as characteristic examples of residual time series. Residuals for the other pulsars can be found in Appendix~C of \citet{agazie2023nanograv}.
More details about the NG15yr dataset can be found in \citet{agazie2023nanograv}.

\section{Results}
\label{sec:results}

\subsection{PSR~J1641+8049 Superior Conjunction Inconsistency Result}
\label{sec:results_conjunction}

Using the optical lightcurves of the PSR~J1641+8049 system, we identify two times of superior conjunction corresponding to peak light. 
% We identify two corresponding points in the light curves: $T_{\rm o, s} = \{58640.0974(15)\,, 58640.1883(15)\}$\,MJD. 
These are $T_{\rm o, s} = \{58640.0974(15)\,, 58640.1883(15)\}$\,MJD. Note the binary period is $0.09087395384(1)$\,day. 
The in-system Romer delay is smaller than the uncertainties, so no correction is needed for $T_{\rm o, s}$.  
The corresponding times of superior conjunction as predicted by radio pulsar timing are $T_{\rm r, s} = \{58640.09735997(5) ,\, 58640.18823392(5)\}$\,MJD. Here, we use the $T_{\rm r,s} = T_{\rm asc} - P_{\rm b}/4$ equation and subtracted integer multiples of $P_{\rm b}$ so that $T_{\rm r,s}$ is as close as possible to $T_{\rm o, s}$. The results are visually summarized in Figure~\ref{fig:optical_light_curve}.

% These different times of superior conjunction give an average time difference of $\Delta t = 4.6 \pm 94$\,s, which corresponds to $\Delta m_{\rm T} = 66\pm1400\,$M$_\oplus$ through Equation~\ref{eq:sup_conj}. 
We find no statistically significant difference in the TSCs, and constrain $|\Delta t|$ to be no higher than 190\,s (95\% confidence).
Therefore, through Equation~\ref{eq:sup_conj}, $|\Delta m_{\rm T}|$ can no higher than 8\,M$_{\rm J}$ between the triangular Lagrange points. Comparing times of superior conjunction also confirms, within the uncertainties, that the companion of PSR~J1641+8049 is being heated symmetrically (see \citealt{kandel2020atmospheric_heating, voisin2020heating_model} for discussions on asymmetric black widow companion heating). Combining both a non-zero $\Delta m$ and asymmetric heating in such a way that their effects on the time of superior conjunction cancel is also possible.

\subsection{Librations Results}
\label{sec:results_librations}
\begin{figure}
        \centering
	\includegraphics[width=\columnwidth]{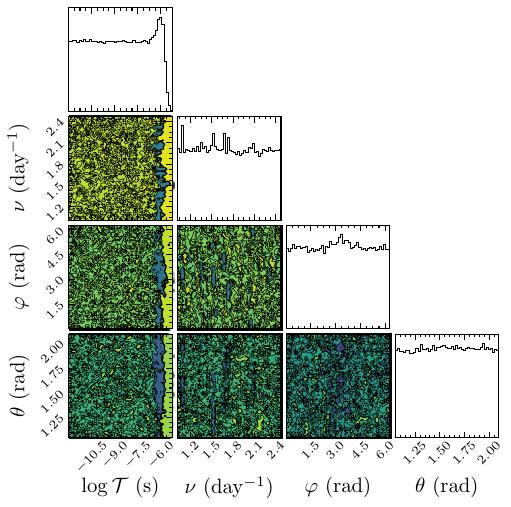}
    \caption{The posterior space of the PSR~J0610$-$2100 libration analysis. Each histogram represents the marginalized posterior of the relevant parameter. Each contour plot represents the joint-marginalized posterior of the two relevant parameters. This is a clear non-detection with almost flat posteriors across all parameters. There is a slight preference for $\log{(\mathcal{T}/\rm{s})} \sim 10^{-6}$, which is similar to the noise floor of this pulsar \citep[][]{agazie2023nanograv}, though $\log{(\mathcal{T}/\rm{s})}$ is agnostic to the other three parameters. Furthermore, the Bayes factor for a non-zero $\log{(\mathcal{T}/\rm{s})}$ is less than 1, so no librations ($\mathcal{T} \equiv 0 $) are probabilistically preferred.}
    \label{fig:J0610_post}
\end{figure}

% With the exception of PSR~J0023+0923, all of the NG15yr pulsars analyzed, we find no evidence of exotrojans in. 
With the exceptions of PSRs~J0023+0923 and J1705$-$1903, we find zero evidence of exotrojans in the NG15yr pulsar systems analyzed.
In Figure~\ref{fig:J0610_post} we showcase a corner plot of the posteriors for PSR~J0610$-$2100, which is characteristic of the other five clear non-detections. To emphasize that these are indeed non-detections, we calculate the Bayes factor, $\mathcal{B_T}$, for each NG15yr pulsar. We use the Savage-Dickey formalism \citep[][]{Dickey1971} adapted for our model,
\begin{equation}
    \mathcal{B_T} = \frac{\rm{prior}(\mathcal{T} = 0)}{\rm{post}(\mathcal{T} = 0)},
\end{equation}
where $\rm{prior}(\mathcal{T} = 0)$ and $\rm{post}(\mathcal{T} = 0)$ are the marginalized prior and posterior probabilities, respectively, when the libration strength is 0 (i.e. Trojans are either entirely absent ($m_{\rm T} \equiv 0$) or not librating ($z\equiv 0$)). Intuitively speaking, this formula describes the relative shift in the probability of a null-detection when TOA data is considered. The libration strength Bayes factor for each NG15yr pulsar is listed in the second column of Table~\ref{tab:upper_limits}.

Upon a positive detection of the signal given by Equation~\ref{eq:libration_residuals}, we employ another Bayes factor,
\begin{equation}
    \mathcal{B}_\xi = \frac{\rm{prior}(\theta = \pi/3)}{\rm{post}(\theta = \pi/3)},
\end{equation}
where $\rm{prior}(\theta = \pi/3)$ and $\rm{post}(\theta = \pi/3)$ are the marginalized prior and posterior probabilities, respectively, when the Trojan mass is zero. 
Therefore, $\mathcal{B_T}$ quantifies the support for a positive detection of our Trojan model, while $\mathcal{B}_\xi$ quantifies the support for the detection being from a massive Trojan, rather than a Trojan with a nonphysical mass. We calculate $\mathcal{B}_\xi$ in $\log{\xi}$ space rather than $\theta$ space (hence the subscript) because $\log{\xi}$ is more sensitive to low Trojan masses. 
The non-zero-Trojan-mass Bayes factors are listed in the last column of Table~\ref{tab:upper_limits} for the two pulsars with positive $\mathcal{T}$ detections.

\begin{table}[ht]
\centering
\begin{tabular}{lcccc}
\hline
PSR & $\mathcal{B_T}$ & $\mathcal{T}_{\rm UL}$ & $m_{\rm T, UL}$ & $\mathcal{B}_\xi$ \\
 &  & ($\mu$s)  & (M$_\oplus$) & \\
\hline
J0023+0923 & $11.5(2)$ & 0.47 & 0.79 & 0.95(7) \\
J0610$-$2100 & $0.824(7)$ & 2.8 & 1.8 & - \\
J0636+5128 & $0.801(8)$ & 0.37 & 0.81 & -\\
J1705$-$1903 & 5700(300) & 1.7 & 2.7 & 0.83(5) \\
J1719$-$1438 & $0.73(1)$ & 1.4 & 2.0 & - \\
J1745+1017 & $0.792(8)$ & 4.2 & 1.5 & - \\
J2214+3000 & $0.90(1)$ & 1.8 & 1.1 & - \\
J2234+0944 & $0.96(2)$ & 1.0 & 0.66& - \\

\hline
\end{tabular}
\caption{Values of $\mathcal{T}_{\rm UL}$, Bayes factor $\mathcal{B_T}$ (with 68\% uncertainty), $m_{\mathrm{T, UL}}$, and $\mathcal{B}_\xi$ for each pulsar. The uncertainty on the Bayes factor is due to the method in calculating the Bayes factor with non-infinite posterior chains. The ULs represent a 95\% credibility. Every NG15yr pulsar except PSRs~J0023+0923 and J1705$-$1903 has $\mathcal{B_T} < 1$. We do not calculate $\mathcal{B}_\xi$ for these six pulsars. The upper limits placed on PSRs~J0023+0923 and J1705$-$1903 were calculated using the same routine as the other pulsars and are independent of the analysis used in investigating a potential detection in the system in Section~\ref{sec:J0023_librations}.}
\label{tab:upper_limits}
\end{table}

Due to mostly non-detections, we place upper limits (ULs) on the libration strength ($\mathcal{T_{\rm UL}}$) and make additional assumptions on the pulsar and pulsar Trojan populations to map the marginalized posterior on $\mathcal{T}$ to a Trojan mass UL ($m_{\rm T, UL}$; see Appendix~\ref{app:upper_limits} for details on the $\mathcal{T} \mapsto m_{\rm T, UL}$ mapping). We choose our ULs to represent 95\% credibility. 
Following \citet{loki2025ISO} we put $\mathcal{T}$ on a linear scale in order to correctly place an upper limit. This requires that we weight the $\mathcal{T}$ histogram bins by a factor of $\mathcal{T}$ to account for the Jacobian in converting between the $\log \mathcal{T}$ and $\mathcal{T}$ probability distributions. \citet{hourihane2023accurate} show that using a linear scale for $\mathcal{T}$ and weighting in this fashion ensure the upper limits do not depend on the $\mathcal{T}$ prior lower-bound. We place the values of $\mathcal{T_{\rm UL}}$ and $m_{\rm T}$ for each pulsar in third and fourth columns, respectively, of Table~\ref{tab:upper_limits}.
% As in \citet{loki2025ISO}, we reweight the posterior on $\mathcal{T}$ to match the posterior expected when instead choosing a uniform, rather than log-uniform, prior on $\mathcal{T}$. 
% Each $\mathcal{T_{\rm UL}}$ and $m_{\rm T, UL}$ is listed in the third and fourth columns, respectively, of Table~\ref{tab:upper_limits}.

Despite positive detections of Equation~\ref{eq:libration_residuals} in the PSR~J0023+0923 and J1705$-$1903 systems, we also place upper limits on them due to the fact that we suspect these are false-positives (see the following subsection). 
In doing so, we assume the signals are not from true librations, and therefore, we follow the same procedure as for the other pulsars. Thus, the Trojan mass upper limits on these two systems are inconsistent with the Trojan masses as calculated in Table~\ref{tab:J0023_J1705_params} because the upper limits assume a certain distribution in $z$, while the results in Table~\ref{tab:J0023_J1705_params} can determine $z$ from the libration parameters. See Section~\ref{sec:J0023_librations} and Appendix~\ref{app:upper_limits} for more details.

\subsection{PSRs~J0023+0923 and J1705$-$1903 Libration Results}
\label{sec:J0023_librations}

\begin{figure}
        \centering
	\includegraphics[width=\columnwidth]{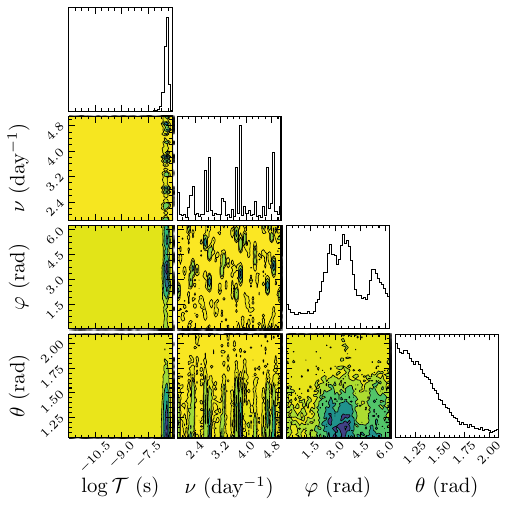}
    \caption{The same as Figure~\ref{fig:J0610_post} except for PSR~J0023+0923. With a Bayes factor of 11.5(2), we have a tentative detection of the signal in Equation~\ref{eq:libration_residuals}. However, we find that the origin of this signal is unlikely to be caused by a Trojan due to a nonphysical Gaussian-average value for $\theta$. See the text for more details.}
    \label{fig:J0023_post}
\end{figure}

\begin{figure}
        \centering
	\includegraphics[width=\columnwidth]{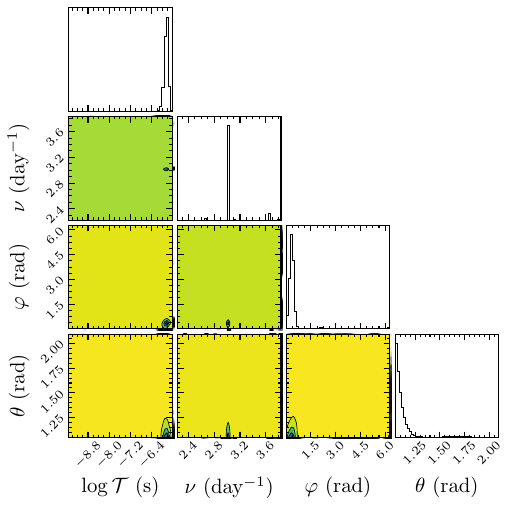}
    \caption{The same as Figure~\ref{fig:J0610_post} except for PSR~J1705$-$1903.}
    \label{fig:J1705_post}
\end{figure}

We find weak evidence of exotrojans in the PSR~J0023+0923 and J1705$-$1903 systems and showcase their posterior corner plots in Figures~\ref{fig:J0023_post} and \ref{fig:J1705_post}. 
% The maximum a posterior (MAP) values (the values that maximize the likelihood function) of the libration parameters are $\log{(\mathcal{T}/\rm{s})} = -6.37$, $\nu = 3.82\,\rm{day}^{-1}$, $\varphi = 5.00\,\rm{rad}$, and $\theta = 1.05$\,rad, each given to an arbitrary three significant figures. 
For PSR~J0023+0923, the posterior space is clearly multimodal, with evidence for sidereal-day-aliasing in $\nu$. Aliasing is an observational cadence effect due to the Nyquist-Shannon sampling theorem, where spurious frequencies can fit the data somewhat well in place of the true frequency. In our case, these different frequencies vary from the true frequency by integer multiples of $f_{\rm sidereal} \approx 1 \,\rm{day}^{-1}$, because the NG15yr TOAs of PSR~J0023+0923 were measured by the Arecibo Observatory, a transit instrument. As for all transit instruments, the Arecibo Observatory had limited pointing capabilities and required the sky to rotate over its field of view. Therefore, every TOA of PSR~J0023+0923 was observed at roughly the same sidereal time. As mentioned in Section~\ref{sec:radio_pulsar_timing_data}, the Nyquist-Shannon sampling theorem is weakened by the vast number of observations occurring at different librational phases, which is why one frequency, $\sim 3.82\,\rm{day}^{-1}$, stands out above the rest. 
PSR~J1705$-$1903 only shows weak aliasing, because it was observed with the Green Bank Telescope, which has full pointing capabilities.
Another feature of both posterior spaces is that $\theta$ tends towards $\pi/3$, which corresponds to $\delta = 0$. Therefore, any Trojans in these two systems will be in the $L_5$ Lagrange point, and thus $m_{\rm c}$ and $m_{\rm T}$ should be exchanged in Figure~\ref{fig:libration_diagram}.

% \begin{figure}
%         \centering
% 	\includegraphics[width=\columnwidth]{J1705_aug28_posteriors_resized.pdf}
% 	% \includegraphics[width=\textwidth]{J0023_aug3_posteriors.pdf}
%     \caption{The same as Figure~\ref{fig:J0023_post} except for PSR~J1705$-$1903. resized}
%     % \label{fig:J1705_post}
% \end{figure}

\begin{figure}
        \centering
	\includegraphics[width=\columnwidth]{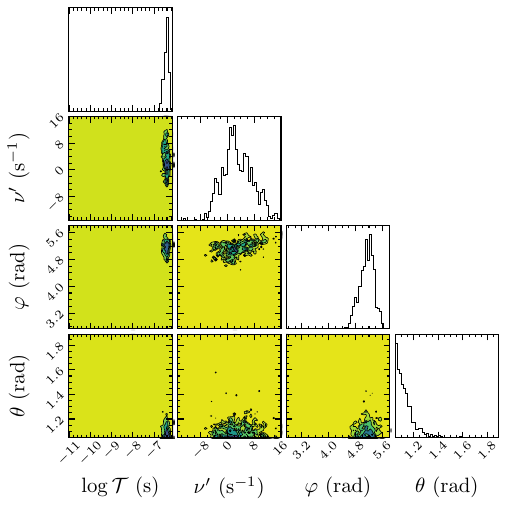}
    \caption{The same as Figure~\ref{fig:J0023_post} (PSR~J0023+0923) except the posterior is restricted to the $\pm$ 0.05\% neighborhood of $\nu = 3.81759\,\rm{day}^{-1}$. The $\pm0.05$\% width was chosen so that several standard deviations were included. A Fisher information analysis reveals this width to be the expected uncertainty in $\nu$. We define $\nu ' = \nu - 3.81759\,\rm{day}^{-1}$. As stated in the text, we find it unlikely that this signal is due to a Trojan.}
    \label{fig:J0023_post_zoomed}
\end{figure}

\begin{figure}
        \centering
	\includegraphics[width=\columnwidth]{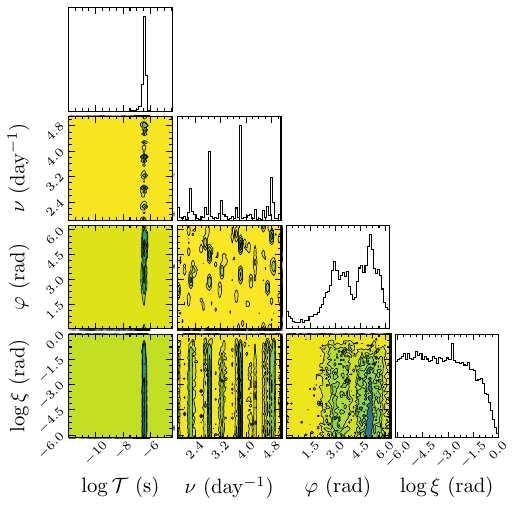}
    \caption{The same as Figure~\ref{fig:J0023_post} (PSR~J0023+0923) except $\log{\xi}$ is searched over uniformly.}
    \label{fig:J0023_post_targeted}
\end{figure}

In Figure~\ref{fig:J0023_post_zoomed} we show the PSR~J0023+0923 posterior space in the $\pm$ 0.05\% neighborhood of $\nu = 3.81759\,\rm{day}^{-1}$. We find that the $\theta$ posterior is fully consistent with that of a Gaussian tail, where the Gaussian is centered at $\theta \approx 0.64$, implying $\delta \approx -0.66$. Furthermore, we find a Gaussian in $\theta$ centered at the same value when searching for a Trojan signal when using a $\theta \sim \mathcal{U}(0,\, 2\pi)$ prior, rather than the physically valid $\theta \sim \mathcal{U}(\pi/3,\, 2\pi/3)$ prior. 
We included a search over this non-physical prior in order to confirm that the source of the posterior is in the non-physical parameter space.
We also search in $\log{\xi}$ space (see Section~\ref{sec:priors}) and find no preferred $\xi$ (see Figure~\ref{fig:J0023_post_targeted}). In fact, we find $\mathcal{B}_\xi < 1$ for both pulsars. We therefore conclude that the signals associated with the $\nu \approx 3.82\,\rm{day}^{-1}$ and $3.02\,\rm{day}^{-1}$ libration frequencies for PSR~J0023+0923 and J1705$-$1903, respectively, are unlikely to be caused by Trojans (see Section~\ref{sec:discussion} for a discussion on possible other causes). However, for instructive purposes, we carry out the following calculations as if the signals were from a Trojan in each system. 

% TODO: update this paragraph for J1705
In this paragraph, we only use samples where $\nu$ is in the $\pm$ 0.05\% neighborhood of $\nu = 3.81759\,\rm{day}^{-1}$ for PSR~J0023+0923 and $3.01600\,\rm{day}^{-1}$ for PSR~J1705$-$1903.
Using Equation~\ref{eq:delta} on the $\theta$ samples and Equations~\ref{eq:nu_cornish} and \ref{eq:m_c_m_p_kappa} on the $\nu$ samples, we find $\delta$ and $q \equiv m_{\rm p} / m_{\rm c}$, placing them in the second and third columns, respectively, of Table~\ref{tab:J0023_J1705_params}. We then use Equation~\ref{eq:z} to calculate $z$, which we include in the fourth column of Table~\ref{tab:J0023_J1705_params}. Exchanging $m_{\rm c}$ and $m_{\rm T}$ in the definition of $\delta$ so that $m_{\rm c}$ is the more massive body, and assuming the pulsar mass function of \citet{antoniadis2016millisecond} (as in Appendix~\ref{app:upper_limits}), we find $m_{\rm T}$, $m_{\rm c}$, and $I$, placing them in the remaining columns of Table~\ref{tab:J0023_J1705_params}. Using an optical light curve analysis of the PSR~J0023+0923 system, \citet{mata2023black0023} find $q = 59_{-11} ^ {+10}$ and $I = 42_{-3}^{+4}$\textdegree{}. The inconsistency of their measurements with ours is another indication that our detected signal is not from a Trojan.

\begin{table*}[ht]
\centering
\begin{tabular}{lcccccc}
\hline
PSR & $\delta$ & $q \equiv m_{\rm p} / m_{\rm c}$ & $z$ (\textdegree) & $m_{\rm T}$ (M$_{\rm J}$) & $m_{\rm c}$ (M$_\odot$) & $I$ (\textdegree) \\
\hline
J0023+0923   & $0.064_{-0.049}^{+0.082}$ & $31.3940_{-0.0006}^{+0.0004}$ & $0.0095_{-0.0054}^{+0.0335}$ & $3.5_{-2.7}^{+5.6}$ & $0.0458_{-0.002}^{+0.014}$ & $13_{-5}^{+1}$ \\
J1705$-$1903 & $0.042_{-0.031}^{+0.061}$ & $29.6472 \pm 0.0005$ & $0.019_{-0.011}^{+0.052}$ & $2.4_{-1.8}^{+4.0}$ & $0.048_{-0.002}^{+0.014}$ & $32_{-12}^{+3}$ \\
\hline
\end{tabular}
\caption{Libration parameters for PSRs~J0023+0923 and J1705$-$1903 if their detected signals are from individual Trojans. Values are derived by exchanging $m_{\rm c}$ and $m_{\rm T}$ in the definition of $\delta$ and assuming the pulsar mass function of \citet{antoniadis2016millisecond}.}
\label{tab:J0023_J1705_params}
\end{table*}

When we isolate the PSR~J0023+0923 posterior space around other peaked libration frequency values, we do find some cases where the marginalized $\theta$ distribution has Gaussian centers in the physically-acceptable $\pi/3 \leq \theta \leq 2\pi/3$ interval. While Trojans in the system offer an explanation for these particular libration frequencies, we caution against this conclusion of a Trojan swarm due to the fact that the most-favored $\nu$ is likely not due to a Trojan.

\subsection{Librating Trojan Prevalence}
\label{sec:eta_trojan}
Accepting the conclusion that we found no librating Trojans in the eight inspected systems, we place a 95\% upper bound on the prevalence of Trojans that are detectable using the libration method, $\eta_{\rm T, Lib}^{\rm UL}$. Here, we ignore the clear differences across the systems, such as orbital parameters and available TOAs. While librations become easier to detect with an increasing number of TOAs, our estimate of $\eta_{\rm T, Lib}^{\rm UL}$ is meant as a first-order, approximate constraint. 

By Bayes' Theorem
\begin{equation}
    p(\eta_{\rm T, Lib} | \text{8 non-detections}) \propto \mathcal{L}(\eta_{\rm T, Lib}) \cdot p(\eta_{\rm T, Lib}),
\end{equation}
where $p(\eta_{\rm T, Lib} | \text{8 non-detections})$ is the probability that $\eta_{\rm T, Lib}$ is the true prevalence of librationally detectable Trojans, given eight systems indeed lacked librationally detectable Trojans; $\mathcal{L}(\eta_{\rm T, Lib})$ is the likelihood of $\eta_{\rm T, Lib}$ and is equal to $p(\text{8 non-detections} | \eta_{\rm T, Lib})$, the probability of eight non-detections given $\eta_{\rm T, Lib}$; and $p(\eta_{\rm T, Lib})$ is the prior probability on $\eta_{\rm T, Lib}$. Choosing a flat prior ($p(\eta_{\rm T, Lib}) \propto 1$) and noting that $p(\text{8 non-detections} | \eta_{\rm T, Lib}) = (1 - \eta_{\rm T, Lib})^8$, we have
\begin{equation}
    p(\eta_{\rm T, Lib}^{\rm UL} | \text{8 non-detections}) = 
    \frac{(1 - \eta_{\rm T, Lib})^8}{\int_0^{1}(1 - \eta_{\rm T, Lib})^8\, {\rm d}\eta_{\rm T, Lib}}.
\end{equation}
Therefore, the 95\% UL is given by solving 
\begin{equation}
     \frac{\int_0^{\eta_{\rm T, Lib}^{\rm UL}}(1 - \eta_{\rm T, Lib})^8 \, {\rm d}\eta_{\rm T, Lib}}{\int_0^{1}(1 - \eta_{\rm T, Lib})^8\, {\rm d}\eta_{\rm T, Lib}} = 0.95 
\end{equation}
for $\eta_{\rm T, Lib}^{\rm UL}$. We find $\eta_{\rm T, Lib}^{\rm UL} = 1- 0.05^{1/9} \approx 28\%$. This tells us that there is a 95\% chance the prevalence of librationally detectable Trojans is no higher than 28\%.

In general, for $N$ non-detections, $\eta_{\rm T, Lib}^{\rm UL}(N) = 1 - 0.05^{1/(N+1)}$. We have $\eta_{\rm T, Lib}^{\rm UL}(N) = |\ln{(0.05)}|/N + \mathcal{O}(1/N^2) \approx3/N$. Therefore, $\eta_{\rm T, Lib}^{\rm UL}(N)$ scales like $1/N$, so more non-detections constraint $\eta_{\rm T, Lib}$ further, but at a decreasing rate.

\section{Discussion and Conclusions}
\label{sec:discussion}
% can do this on more lightcurves. also novel way to check how symmetrical the heating models are
% can do this on the 20yr dataset --- part. inter. for J0023
% can do this on other IPTA datasets
% CHIME (and slow?) pulsars maybe
% broader impacts: the disk environment may be hostile to Trojans. etc.... Talk about eta_troj here

This study presents the first search for pulsar-bound exotrojans. Trojans in general are difficult to detect through methods that rely exclusively on the reflex motion of their host star because these effects are degenerate with those of the dominant co-orbital. Therefore, standard pulsar exoplanet searches cannot probe for co-orbital bodies. Trojans instead require explicit searches, due to their unique, subtle effects on the pulse times of arrival. A librating Trojan detection would improve constraints placed on the secondary mass and orbital inclination, giving good motivation for targeted Trojan searches.

We used two distinct methods to place upper mass limits on Trojans in eight pulsar binary systems. One method combines radio pulsar timing with optical-band light curves in order to determine the difference in location between the fictitious combined companion and the actual companion. Upon finding no significant difference, we concluded that the mass difference between the triangular Lagrange points can be no more than 8\,M$_{\rm J}$ in the PSR~J1641+8049 system. Our use of this method reaffirms the symmetry of black widow heating models. The other method requires no optical data but instead relies on the Trojan and secondary librating with respect to their stable configuration. Despite the highly precise pulse times of arrival in the NG15yr narrowband dataset, we find only weak evidence of librations in only two of the eight NG15yr systems studied here. We place $\sim 1 \,\rm{M}_\oplus$ upper limits on the eight NG15yr systems assuming some properties of the pulsar population as well the hypothetical pulsar Trojan population (see Appendix~\ref{app:upper_limits}). 

While we find signals consistent with librations in the PSR~J0023+0923 and PSR~J1703$-$1903 systems, the signals are likely associated with a nonphysical part of the parameter space whose uncertainties spread into the physically acceptable space. However, if the signals are rather from Trojans, we find $m_{\rm T} \sim 2$--4$\,\rm{M}_{\rm J}$. Other determined properties, like $m_{\rm c}$ and $I$, of the PSR~J0023+0923 system are at odds with \citet{mata2023black0023}, weakening any conclusion that the detected signal is Trojan-caused. Furthermore, $\mathcal{B}_\xi < 1$ for both pulsar systems, meaning the detections of the Equation~\ref{eq:libration_residuals} timing model are almost certainly false positives with respect to Trojan detection. We suspect that the true origin is due to complex and chaotic mass transfers in black widow systems, due to the fact that PSR~J0023+0923 has known orbital variability \citep[][]{bak2020timing_J0023_orbital_variability, agazie2023nanograv} and PSR~J1703$-$1903 experiences radio eclipses for roughly 30\% of its orbit \citep[][]{agazie2023nanograv}. Indeed, the timing of both of these pulsars requires five binary frequency time derivatives 
($\mathrm{d}n_{\mathrm b}/\mathrm{d}t, \,\mathrm{d}^2n_{\mathrm b}/\mathrm{d}t^2, \, ..., \, \mathrm{d}^5n_{\mathrm b}/\mathrm{d}t^5$; \citealt{agazie2023nanograv}). We also identify statistically significant red noise in PSR~J1703$-$1903. Additionally, \citet{agazie2023nanograv} find that the parallax determination of PSR~J0023+0923 is covariant with the binary frequency derivatives, further complicating its stable timing. That being said, as to why a model with two frequencies straddling the orbital frequency by the same margin ($\nu$; see Equation~\ref{eq:libration_residuals_alt}) is preferred over a simpler type of ``noise" is unclear.

% TODO: add some comments about the status of these two BWs in the array.

% In any case, with a possible detection ($\mathcal{B_T} = 11.5(2)$) of the timing signal modeled by Equation~\ref{eq:libration_residuals} in the case of PSR~J0023+0923 and a significant detection ($\mathcal{B_T} = 5700(300)$) of the same for PSR~J1705$-$1903, 
% In any case, with the timing residuals of PSR~J0023+0923 possibly favoring ($\mathcal{B_T} = 11.5(2)$) Equation~\ref{eq:libration_residuals} being included with the original timing model as compared to a timing model without Equation~\ref{eq:libration_residuals}, and a significant preference ($\mathcal{B_T} = 5700(300)$) of the same for PSR~J1705$-$1903, the status of these two pulsars in the NANOGrav pulsar timing array may be called into question. 
In any case, the timing residuals of PSR~J0023+0923 possibly favor ($\mathcal{B_T} = 11.5(2)$) Equation~\ref{eq:libration_residuals} being included with the original timing model, and the residuals significantly prefer ($\mathcal{B_T} = 5700(300)$) of the same for PSR~J1705$-$1903. Therefore, the status of these two pulsars in future pulsar timing array datasets may be called into question. 
Notably, the MeerKAT Pulsar Timing Array has already removed PSR~J1705$-$1903 from its array in its most recent data release \citep{miles2025meerkat}.
While \citet{bochenek2015feasibility} show that including up to four orbital frequency derivatives does not affect a pulsar's sensitivity to gravitational waves, the Trojan libration model has not been applied to any pulsar before this work. If these pulsars are to be included in future pulsar timing array datasets, such as the forthcoming NANOGrav 20-year dataset, more noise modeling or analysis may be required, which may weaken the effectiveness of these pulsars in detecting the gravitational wave background. Conversely, including this new source of ``noise" may improve their effectiveness as compared to where the pulsars currently stand.

There are several clear paths to continuing the pulsar Trojan search. There exist myriad light curves of pulsar binary systems, all of which can have their times of superior conjunction compared to those predicted by pulsar timing. Furthermore, if several light curves are obtained on the same pulsar with the explicit goal of lowering the uncertainty in the time of superior conjunction, libration-independent constraints on the pulsar Trojan population will lower in tandem. In terms of the libration method, plenty of TOA data sets are available for analysis, with several high-quality ones in the making. Currently available high-cadence datasets include those from the the European Pulsar Timing Array \citep{antoniadis2023epta_second}, the MeerKAT Pulsar Timing Array \citep{miles2025meerkat}, and the Parkes Pulsar Timing Array \citep{zic2023parkes_ppta}. 
Gamma-ray pulsar timing arrays \citep[e.g.,][]{fermi2022gamma, kerr2024upgrading} present yet another way to search for Trojan librations.
Because this work was done on the NG15yr dataset, the NANOGrav 20-year dataset will allow for the quickest updates on the Trojan mass constraints or may even reveal a detection. 

Our analysis reveals that pulsar Trojans detectable through their librations are likely somewhat rare, with there being no more than one for every three pulsars that satisfy Equation~\ref{eq:gascheau_stability}. One explanation is that pulsar Trojans are strongly damped into their stable configuration point, lowering the libration amplitude. This seems plausible since pulsar Trojans would have to form from a dense debris or accretion disk. Alternatively, noting that $\mathcal{T} \propto z \cdot m_{\rm T}$, the pulsar Trojan masses may simply be too low to detect for their given distribution in libration amplitude. Another explanation is that the conditions amenable to Trojan formation are also poor environs for those same Trojans. Mass transfer and disk dynamics in pulsar binary systems are clearly turbulent, and so any interloping Trojan may find itself ejected from the system or torn apart. Lastly, conditions in pulsar binaries may never be friendly to Trojan formation to begin with. Future pulsar Trojan searches, as discussed above, will aid in lowering the upper limit on the librating pulsar Trojan prevalence, and will also constrain the non-librating pulsar Trojan occurrence rate through the superior conjunction method.

Our work here not only represents the first search for Trojans in pulsar binary systems, but also, along with \citet{loki2025ISO}, showcases a rigorous and thorough way to analyze deterministic signals in pulse TOAs. The search procedure can be extended to ordinary pulsar exoplanets, pulsar exomoons (or, equivalently, in terms of TOA residuals, planets in the S-type orbital configuration), and even massive bodies in our own Solar System. Such a Solar System body would be correlated across all pulsars, because the location of the Solar System barycenter would shift unexpectedly in time. This is the so-called ``Earth term" and can be used to, for example, detect unknown massive Trojans of any of the planets, discover unknown primordial black holes that orbit the sun, or constrain the mass of the known objects in the Solar System. 

\section*{Data availability}
The NG15yr narrowband data set is available at \url{https://zenodo.org/records/14773896} \citep{the_nanograv_collaboration_2025_14773896}.

\section*{Author Contributions}
J.D.T. led the project, extended the libration theory of \citet{leleu_mainpaper_2015} as relevant for black widow systems, produced the figures, conducted the analysis, and prepared the text with input from E.F., L.D., S.Z., A.Y.K., and J.G. 
L.D. provided the original code infrastructure for the use of \texttt{ENTERPRISE} and \texttt{PTMCMC} before the FFO-to-Trojan transition and offered key insights into the Bayesian inference used. 
S.Z. and A.Y.K. carried out the optical-band light curve analysis in determining the optical times of superior conjunction and wrote the corresponding Section~\ref{sec:optical_data}. 
A.Y.K. provided the optical light curve data and contributed useful discussions regarding PSR~J1641+8049. 
J.G. provided essential high-performance computing support.
J.G., P.B.D., M.K., D.J.N., D.L.K., and T.J.W.L. contributed insightful discussions.
T.J.W.L. and D.R.L. gave helpful comments. 
E.F., J.G., P.B.D., M.K., D.J.N., D.L.K., D.R.L., and the remaining authors contributed to the collection and analysis of the NANOGrav 15\,yr dataset.

\begin{acknowledgments}

We thank Reynier Squillace and the anonymous reviewer for helpful comments. 
Computational resources were provided by the Link HPC cluster and cyber-infrastructure, which is maintained by the Center for Gravitational Waves and Cosmology at West Virginia University and funded in part by NSF IIA-1458952 \& NSF PHY-2020265. 
Additional computational resources were provided by the WVU Research Computing Thorny Flat HPC cluster, partly funded by NSF OAC-1726534. 
This research was made
possible by the NASA West Virginia Space Grant Consortium, Grant~\#~80NSSC25M7079, which provided individual support to J.D.T. 
This work has been carried out as part of the NANOGrav collaboration, which receives support from the National Science Foundation (NSF) Physics Frontiers Center award numbers 1430284 and 2020265. 
The Arecibo Observatory is a facility of the NSF operated under a cooperative agreement (No. AST-1744119) by the University of Central Florida (UCF) in alliance with Universidad Ana G. M{\'e}ndez (UAGM) and Yang Enterprises (YEI), Inc. 
The Green Bank Observatory is a facility of the NSF operated under a cooperative agreement by Associated Universities, Inc. 
The National Radio Astronomy Observatory is a facility of the NSF operated under a cooperative agreement by Associated Universities, Inc. P.R.B.\ is supported by the Science and Technology Facilities Council, grant number ST/W000946/1.
P.R.B.\ is supported by the Science and Technology Facilities Council, grant number ST/W000946/1. % start of nanoauthor acknowledgments
H.T.C.\ acknowledges funding from the U.S. Naval Research Laboratory.
Pulsar research at UBC is supported by an NSERC Discovery Grant and by CIFAR.
K.C.\ is supported by a UBC Four Year Fellowship (6456).
M.E.D.\ acknowledges support from the Naval Research Laboratory by NASA under contract S-15633Y.
T.D.\ and M.T.L.\ received support by an NSF Astronomy and Astrophysics Grant (AAG) award number 2009468 during this work.
E.C.F.\ is supported by NASA under award number 80GSFC24M0006.
% G.E.F.\ is supported by NSF award PHY-2011772.
D.C.G.\ is supported by NSF Astronomy and Astrophysics Grant (AAG) award \#2406919.
D.R.L.\ and M.A.M.\ are supported by NSF \#1458952.
M.A.M.\ is supported by NSF \#2009425.
The Dunlap Institute is funded by an endowment established by the David Dunlap family and the University of Toronto.
T.T.P.\ acknowledges support from the Extragalactic Astrophysics Research Group at E\"{o}tv\"{o}s Lor\'{a}nd University, funded by the E\"{o}tv\"{o}s Lor\'{a}nd Research Network (ELKH), which was used during the development of this research.
H.A.R.\ is supported by NSF Partnerships for Research and Education in Physics (PREP) award No.\ 2216793.
S.M.R.\ and I.H.S.\ are CIFAR Fellows.
Portions of this work performed at NRL were supported by ONR 6.1 basic research funding.
S.Z. and A.K. thank Daniel Mata Sanchez for providing the optical data. 
SZ acknowledges the DGAPA-PAPIIT grant IN119323. 
AK acknowledges the DGAPA-PAPIIT grant IA105024.
% Pulsar research at UBC is supported by an NSERC Discovery Grant and by CIFAR.
% K.C.\ is supported by a UBC Four Year Fellowship (6456).
% M.E.D.\ acknowledges support from the Naval Research Laboratory by NASA under contract S-15633Y.
% T.D.\ and M.T.L.\ received support by an NSF Astronomy and Astrophysics Grant (AAG) award number 2009468 during this work.
% E.C.F.\ is supported by NASA under award number 80GSFC24M0006.
% G.E.F.\ is supported by NSF award PHY-2011772.
% D.R.L.\ and M.A.M.\ are supported by NSF \#1458952.
% M.A.M.\ is supported by NSF \#2009425.
% The Dunlap Institute is funded by an endowment established by the David Dunlap family and the University of Toronto.
% T.T.P.\ acknowledges support from the Extragalactic Astrophysics Research Group at E\"{o}tv\"{o}s Lor\'{a}nd University, funded by the E\"{o}tv\"{o}s Lor\'{a}nd Research Network (ELKH), which was used during the development of this research.
% H.A.R.\ is supported by NSF Partnerships for Research and Education in Physics (PREP) award No.\ 2216793.
% S.M.R.\ and I.H.S.\ are CIFAR Fellows.
% Portions of this work performed at NRL were supported by ONR 6.1 basic research funding.

\end{acknowledgments}


\vspace{5mm}
\facilities{GTC, CHIME, GBT, Arecibo}


\software{\texttt{PINT} \citep{luo2019ascl, luo2021pint}, \texttt{rebound} \citep{rebound, reboundias15}, \texttt{ENTERPRISE} \citep{EllisVallisneri+2017,JohnsonMeyers+2023},
\texttt{enterprise\_extensions} \citep{TaylorBaker+2021,JohnsonMeyers+2023},
\texttt{PTMCMCSampler} \citep{EllisvanHaasteren2017,JohnsonMeyers+2023},
\texttt{numpy} \citep{HarrisMillman+2020}, 
\texttt{matplotlib} \citep{Hunter2007}, 
\texttt{corner} \citep{Foreman-Mackey2016}.}

\appendix
\section{Effect of Librations on Pulsar Times of Arrival}
\label{app:libration_derivation}
Here, we largely follow the calculations of \citet{leleu_mainpaper_2015} except we include the $\mathcal{O}(\nu / n_{\rm{b}})$ term that cannot be ignored for the pulsar companion masses of interest in our search. Many of the variables are defined in Section~\ref{sec:libration_theory}, but we do repeat some details here for clarity. Referencing Figure~\ref{fig:libration_diagram}, we have the Trojan and companion librating with respect to each other, with the libration angle given by $\zeta = z \cos{(\nu\tilde{t} + \varphi)} + \pi/3$ (Equation~\ref{eq:zeta_approx}). The mass-averaged longitude, $\lambda = n_{\rm b} \tilde{t} + \lambda_0$, follows the mean motion and is unaffected by the librations. The angles of the Trojan and companion with respect to $\lambda$ are $\delta \zeta$ and $(1-\delta) \zeta$, respectively. Therefore, as \citet{robutel2011analytical} showed,
\begin{align}
    \lambda_{\rm T} &= \lambda + \delta \zeta, \\
    \lambda_{\rm c} &= \lambda - (1-\delta) \zeta.
\end{align}
The semi-major axes of the Trojan and the companion in the pulsar-centric reference frame also oscillate with respect to the mass-averaged semi-major axis, $a$, and are given by \citet[][]{robutel2011analytical} as
\begin{align}
    a_{\rm T} &\approx a\left(1 - \frac{2\delta \dot{\zeta}}{3n_{\rm b}}\right),\\
    a_{\rm c} &\approx a\left(1 + \frac{2(1-\delta) \dot{\zeta}}{3n_{\rm b}}\right).
\end{align}
Explicitly, $\dot{\zeta} = z \nu \sin{(\nu \tilde{t} + \varphi})$. 
% Hereafter $C = $ and $S = $.

In the pulsar-centric reference frame,
\begin{align}
    \boldsymbol{r}_{\rm T} &= a_{\rm T} e^{i\delta \zeta} e^\lambda,\\
    \boldsymbol{r}_{\rm T} &= a_{\rm T} e^{i(\delta - 1) \zeta} e^\lambda,
\end{align}
where we use the $\boldsymbol{r} = x + iy$ complex notation. In the barycentric reference frame, 
\begin{equation}
    \boldsymbol{r}_{\rm p} = -\mu[(1-\delta)\boldsymbol{r}_{\rm T} + \delta \boldsymbol{r}_{\rm c}].
\end{equation}

In the $z \ll 1$ limit, we find that $\boldsymbol{r}_{\rm p}$ can be separated into two parts, 
\begin{equation}
    \boldsymbol{r}_{\rm p} = -\boldsymbol{r}_{\rm f} - \boldsymbol{r}_{\rm L},
\end{equation}
where
\begin{equation}
    \boldsymbol{r}_{\rm f} = \mu a (1-\delta + \delta e^{-i\pi/3})e^{i(\lambda + \delta \pi/3)}
\end{equation}
is the term that is completely degenerate with that of a single companion and 
\begin{align}
    % \begin{equation}
    \boldsymbol{r}_{\rm L} =& \mu a z \delta (1-\delta)\left( i\cos{(\nu \tilde{t} + \varphi)}+\frac{2\nu}{3n_{\rm b}}\sin{(\nu \tilde{t} + \varphi)}\right) \nonumber\\
    & \times e^{i\pi/3} e^{i(\lambda + \delta \pi/3)} + \mathcal{O}\left(\frac{\nu}{n_{\rm b}}z^2\right)
    % \end{equation}
\end{align}
represents non-degenerate terms due to librations. 
By convention, we place the Earth such that it lies in the $\hat{x}\hat{z}$-plane (see Figure~\ref{fig:libration_diagram}) and makes an angle $I$ with the $\hat{z}$-axis. Therefore, by Equation~\ref{eq:ross_toa_delay},
\begin{equation}
    \label{eq:tau_full}
    \Delta \tau _{\rm{full}} = \Re(\boldsymbol{r}_{\rm f} + \boldsymbol{r}_{\rm L})\frac{\sin{I}}{c},
\end{equation}
where $\Delta \tau _{\rm{full}}$ denotes the full R{\o}mer delay due to the Trojan and companion. The final $\Delta \tau$ does not include terms degenerate with the timing model.

For the degenerate term, we have 
% \begin{equation}
\begin{align}
    \Re(\boldsymbol{r}_{\rm f})\frac{\sin{I}}{c} =& \frac{\mu a \sin{I}}{c} \sqrt{1 - \delta (1-\delta)} \nonumber \\
    & \times \cos{\left[n \tilde{t} + \lambda_0 + \delta \frac{\pi}{3} - \arctan{\left(\frac{\delta \sqrt{3}}{2 - \delta}\right)}\right]} .
\end{align}
% \end{equation}
% \begin{align}
%     \Re(\boldsymbol{r}_{\rm f})\frac{\sin{I}}{c} =& \frac{\mu a \sin{I}}{c} \sqrt{1 - \delta (1-\delta)} \\
%     & \times \cos{\left[n \tilde{t} + \lambda_0 + \delta \frac{\pi}{3} - \arctan{\left(\frac{\delta \sqrt{3}}{2 - \delta}\right)}\right]}. \nonumber
% \end{align}
This is already included in the timing model (Section~\ref{sec:bayesian_parameters}), and so we can exploit the fact that $T_{\rm{asc}}$ and $K = \alpha \sqrt{1 - \delta (1-\delta)} = \mu a \sin{I} \sqrt{1 - \delta (1-\delta)}/c$ are already known. Without prior knowledge of a Trojan,
\begin{equation}
    K_2 = \frac{m_{\rm f}}{m_{\rm p} + m_{\rm f}} \frac{a \sin{I}}{c},
\end{equation}
where $K_2$ is the ordinary two-body equivalent to $K$.
Setting $K = K_2$ gives the mass of the combined fictitious object,
% \begin{widetext}
\begin{align}
\label{eq:app_mf}
m_{f} &=\frac{ m_{p} (m_{\rm T} + m_{\rm c}) \sqrt{1- \delta (1 - \delta)} } { m_{p} + (m_{\rm T} + m_{\rm c}) ( 1 - \sqrt{1 - \delta \left(1 - \delta \right)} )} \\
&= (m_{\rm c} + m_{\rm T})\sqrt{1 - \delta (1- \delta)} + \mathcal{O}\left(\frac{m_{\rm T}}{m_{\rm p}},\,\frac{m_{\rm c}}{m_{\rm p}}\right). \nonumber
\end{align}
% \end{widetext}

To zeroth order in the orbital eccentricity, \citet{lange2001precision} defined $T_{\rm{asc}}$ such that $\Delta \tau _{2} = K \sin{(\Phi)} = K \sin{(n_{\rm b} \tilde{t})}$, where $\Delta \tau _{2}$ is the time delay from an ordinary two-body system and $\Phi = n_{\rm b} \tilde{t}$. $\phi = \Phi - \pi/2$ is shown in Figure~\ref{fig:libration_diagram}. The $-\pi/2$ phase difference arises from how \citet{lange2001precision} reference $\Phi$ to the negative $\hat{y}$-axis of our convention. Therefore, 
\begin{equation}
    \phi_0 = -\frac{\pi}{2}.
\end{equation}
% Note that $\phi$ and $\lambda$ are close but not identical, with 
% \begin{equation}
% \phi - \lambda = \frac{\delta \pi}{3}- \arctan{\left( \frac{\delta \sqrt{3}}{2 - \delta} \right)}\in \,\sim [-0.02,\,0.02]. 
% \end{equation}
Matching the phases of $\Delta \tau _{2}$ and $\Re(\boldsymbol{r}_{\rm f})\sin{(I)/c}$ gives 
\begin{equation}
    \label{eq:phase_match}
    \lambda_0 + \delta \frac{\pi}{3} - \arctan{\left(\frac{\delta \sqrt{3}}{2 - \delta}\right)} = -\frac{\pi}{2}.
\end{equation}
Therefore, $\phi$ and $\lambda$ are close but not identical, with 
\begin{equation}
\phi - \lambda = \frac{\delta \pi}{3}- \arctan{\left( \frac{\delta \sqrt{3}}{2 - \delta} \right)}\in \,\sim [-0.02,\,0.02]. 
\end{equation}
Note that $\phi \equiv \lambda$ only when $\delta = 0$, 1/2, or 1.

Continuing Equation~\ref{eq:tau_full} with the non-degenerate term, and ignoring $\mathcal{O}(\nu z^2/n_{\rm b})$ effects, we have 
% \begin{widetext}
\begin{align}
    \label{eq:libration_term}
    \Re(\boldsymbol{r}_{\rm T})\frac{\sin{I}}{c} =& \frac{\mu a \sin{I}}{c} z \delta (1-\delta) \nonumber \\
    & \times \Big[ \cos{(\nu \tilde{t} + \varphi)} \cos{\left(n\tilde{t} + \lambda_0 + \delta \frac{\pi}{3}+\frac{5\pi}{6}\right)} \nonumber \\
    & + \frac{2\nu}{3n_{\rm b}}\sin{(\nu \tilde{t} + \varphi)} \cos{\left(n\tilde{t} + \lambda_0 + \delta \frac{\pi}{3} + \frac{\pi}{3}\right)} \Big].
\end{align}
% \end{widetext}
The overall coefficient defines $\mathcal{T}$ and the phase 
\begin{equation}
    \label{eq:theta_simp}
    \theta = \lambda_0 + \delta \frac{\pi}{3}+\frac{5\pi}{6}
\end{equation} 
is defined for simplification. Combining Equations~\ref{eq:phase_match} and \ref{eq:theta_simp} gives
Equation~\ref{eq:theta}. The pulsar timing residuals that are not degenerate with the timing model are given by Equation~\ref{eq:libration_term} and simplified in Equation~\ref{eq:libration_residuals}. 

We can write Equation~\ref{eq:libration_residuals} in a way that shows the periodicity involved directly, and represents how it would appear in the frequency domain. Namely,
\begin{equation}
    \label{eq:libration_residuals_alt}
    \Delta \tau = \mathcal{T} \,\sum_\pm \left(\frac{1}{2} \mp \frac{1}{3}\frac{\nu}{n_{\rm b}}\right)
\cos{\left[(n_{\rm b}\pm \nu)\tilde{t} + (\theta \pm \varphi)\right]},
\end{equation}
where $\sum_\pm$ denotes the sum over the plus and minus terms (e.g., $\sum_\pm (a\pm b) = (a+b) + (a-b)$). Thus, in frequency space, we expect to see a double peak around $n_{\rm b}$ at $n_{\rm b} \pm \nu$, where the $n_{\rm b} -\nu$ peak is stronger. \citet{leleu_mainpaper_2015} identify this double peak structure in their Figure~12. We conducted three-body simulations in the integrator \texttt{REBOUND} \citep[][]{rebound, reboundias15} and verify that the $n_{\rm b} -\nu$ peak is stronger when $\mathcal{O}(\nu / n_{\rm b})$ term cannot be ignored due to $\mu$ being sufficiently high (see Equation~\ref{eq:nu_leleu}). 

We identify two corrections to the model in \citet{leleu_mainpaper_2015}, even after we remove the $\mathcal{O}(\nu / n_{\rm b})$ term. The first correction stems from their Equation~30, line 2. Here, they list $i(1-\delta)$, which instead should be $i(\delta - 1)$. This is because in their Equations~9 and 10, there is a minus sign in front of the $i(1-\delta)$ exponent. This error leads to an additional factor of $\sqrt{3}$ in their derivation of $S_{1}^* = S_{-1}^*$ (see their Equation~32), which are equivalent to our $\mathcal{T}/2$. We place a superscript asterisk ($*$) on variables in their notation to avoid confusion.

Another error stems from their Equation~11, and has less obvious effects due to differences in convention. Here, there should be an overall minus sign on the right-hand side of their Equation~11 so that the center of mass is at the origin of the barycentric frame, by definition. This error leads to misattribution of their $\phi_{\pm1}^*$, which are equal to our $(\theta \pm \varphi)$. In order to compare our results to theirs, we first map our convention to theirs. The mapping is
\begin{equation}
    (\nu T_{\rm asc} + \varphi, \, n_{\rm b} T_{\rm asc} + \lambda_0) \mapsto (-\nu  t_0^*,\,\lambda _0^*).
\end{equation}
Then we obtain
\begin{equation}
    \phi_{\pm 1}^* = \theta \pm \varphi = \lambda_0^* + \delta \pi/3 + 5\pi/6 \mp \nu t_0^*.
\end{equation}
Their $\phi^*_{\pm 1}$ has a phase addition of $-\pi/6$, rather than $5\pi/6$.
We note that neither of their errors affects the model waveform, but rather the physical implications (such as assigning masses or the inclination to a system with given libration parameters).

% Even when ignoring the $\mathcal{O}(\nu / n_{\rm b})$ term, Equation~\ref{eq:libration_residuals_alt}, and therefore Equation~\ref{eq:libration_residuals}, is a correction of the signal derived in \citet{leleu_mainpaper_2015}. In their definition of $S_1  = S_2 $, which is equal to our $\mathcal{T}/2$, they include an extra factor of $\sqrt{3}$ in their Equation~32. Furthermore, mapping our convention to theirs and placing variables in their convention in brackets to avoid confusion, $(\nu T_{\rm asc} + \varphi, \, n_{\rm b} T_{\rm asc} + \lambda_0) \mapsto (-\nu  [t_0],\,[\lambda _0])$, reveals that our definitions of $\theta$ and $\varphi$ used in Equation~\ref{eq:libration_residuals_alt} correct their Equation~44. Namely, we find $[\phi_{\pm 1}] = \theta \pm \varphi = [\lambda_0] + \delta \pi/3 + 5\pi/6 \mp \nu [t_0]$.

% Even when ignoring the $\mathcal{O}(\nu / n_{\rm b})$ term, Equations~\ref{eq:libration_residuals_alt} (and therefore Equation~\ref{eq:libration_residuals} too) is inconsistent with the model derived in \citet{leleu_mainpaper_2015}. T, we correct it by removing the additional factor of $\sqrt{3}$ in their definition of $S_1 = S_0$ (which are equal to $\mathcal{T}/2$) in their Equation~32. 

\section{Upper Limits on the Trojan Mass}
\label{app:upper_limits}
% derive m_T = Tm_c / (Kz)
% state the MC technique for sampling (the reweighted) T, m_c (from m_p), and z. 
To turn an UL on $\mathcal{T}$ into an upper limit on $m_{\rm T}$, some assumptions have to be made about the population of undetected Trojans in the pulsar binary system, as well as the pulsar and its secondary. Here, we only concern ourselves with the most massive of such Trojans. We start with the definition of $\mathcal{T} = \alpha z \delta (1-\delta)$. Defining $\epsilon = m_{\rm T}/ m_{\rm c}$ and ignoring $\mathcal{O}(\epsilon ^2)$ terms, we find
\begin{align}
    \alpha &= K\left(1+\frac{1}{2}\epsilon\right),\\
    \delta &= 1 - \epsilon.
    % 1-\delta &= \epsilon.
\end{align}
% lay out alpha, delta, and 1 - delta. Then give the resultant T and solve for m_T.
Therefore,
\begin{equation}
    \label{eq:mTupper}
    \mathcal{T} = Kz\epsilon
    \implies m_{\rm T} = \frac{\mathcal{T}m_{\rm c}}{Kz}.
\end{equation}

To find the 95\% UL on $m_{\rm T}$, we sample $\mathcal{T}$, $m_{\rm c}$, $K$, and $z$ from their expected distributions in a Monte Carlo fashion. $\mathcal{T}$ is sampled from the reweighted marginalized posterior (see Section~\ref{sec:results_librations}). $K$ is fixed to the value of $a_{\rm p} \sin{I}$ determined in the timing model because its uncertainty is extremely low compared to the other parameters in Equation~\ref{eq:mTupper}. While pulsar Trojans clearly do not share the same evolutionary history as their Jovian brothers, we choose to sample $z$ from the distribution of Jupiter Trojans with inclinations less than 10\textdegree{} as shown in Figure~3 of \citet{marzari_libration_amplitude_distribution_2003}. $m_{\rm c}$ is sampled indirectly by solving the binary mass function (Equation~\ref{eq:binary_mass_function}) for each sample of $m_{\rm p}$ and $\sin{I}$. We choose the standard $\cos{I} \sim \mathcal{U}(0,\, 1)$, which assumes a random orientation on the unit sphere of the vector orthogonal to the orbital plane of the pulsar and companion. We choose $m_{\rm p}$ from the bimodal pulsar mass function found by \citet{antoniadis2016millisecond}.

\bibliography{main}{}
\bibliographystyle{aasjournal}



\end{document}